\documentclass[11pt,a4paper]{article}
\usepackage[T1]{fontenc}
\usepackage{amsmath,amssymb}
\usepackage{mathptmx}
\usepackage{bm}
\usepackage[a4paper,margin=25mm]{geometry}
\usepackage{graphicx}
\usepackage{booktabs}
\usepackage[separate-uncertainty=true,detect-all]{siunitx}
\usepackage[nomove]{cite}
\usepackage[font=small,labelfont=bf,labelsep=period,justification=justified]{caption}
\usepackage{titlesec}
\usepackage{orcidlink}
\usepackage{xurl}
\usepackage{hyperref}
\hypersetup{colorlinks=true,linkcolor=blue!60!black,citecolor=blue!60!black,urlcolor=blue!60!black,pdftitle={Energy-filtered readout preserves gap-enhanced sensing in interacting quantum dots},pdfauthor={Shirwan Abdullah}}
\usepackage{setspace}
\usepackage{microtype}
\usepackage{lineno}
\newif\ifreviewcopy\reviewcopyfalse   % set \reviewcopytrue for the line-numbered version sent to the referees
\titleformat{\section}{\normalfont\large\bfseries}{\thesection.}{0.6em}{}
\titleformat{\subsection}{\normalfont\normalsize\itshape}{\thesubsection.}{0.6em}{}
\titlespacing*{\section}{0pt}{3.0ex plus 1ex minus .2ex}{1.4ex plus .2ex}
\titlespacing*{\subsection}{0pt}{2.4ex plus 1ex minus .2ex}{1.0ex plus .2ex}

\newcommand{\Fm}{F_{\mathrm{mean}}}
\newcommand{\Xf}{X_{0f}}
\newcommand{\Gmeas}{\Gamma_{\mathrm{meas}}}
\newcommand{\Gup}{\Gamma_{\uparrow}}
\newcommand{\Gdn}{\Gamma_{\downarrow}}
\newcommand{\Asym}{\mathcal{A}}
\newcommand{\Tr}{\mathcal{T}}
\newcommand{\Ha}{\mathrm{Ha}^{*}}
\newcommand{\Eq}[1]{equation~(\ref{#1})}
\newcommand{\Eqs}[2]{equations~(\ref{#1}) and (\ref{#2})}
\newcommand{\Fig}[1]{figure~\ref{#1}}
\newcommand{\Tab}[1]{table~\ref{#1}}
\newcommand{\Sec}[1]{section~\ref{#1}}
\newcommand{\App}[1]{appendix~\ref{#1}}
\newcommand{\APP}[1]{Appendix~\ref{#1}}
\newcommand{\Figs}[2]{figures~\ref{#1} and \ref{#2}}

\newcommand{\FIG}[1]{Figure~\ref{#1}}
\newcommand{\TAB}[1]{Table~\ref{#1}}
\newcommand{\SEC}[1]{Section~\ref{#1}}
\begin{document}
\ifreviewcopy\linenumbers\fi

\begin{flushleft}
{\LARGE\bfseries Energy-filtered readout preserves gap-enhanced sensing in interacting quantum dots\par}
\vspace{5mm}
{\large Shirwan Abdullah\,\orcidlink{0009-0006-5971-8537}\par}
\vspace{2mm}
{\small Kurdistan Institution for Strategic Studies and Scientific Research, Sulaymaniyah 46001, Kurdistan Region, Iraq\par}
\vspace{1mm}
{\small E-mail: \href{mailto:shirwan.abdullah@kissr.edu.krd}{shirwan.abdullah@kissr.edu.krd}\par}
\end{flushleft}

\vspace{2mm}
\noindent\rule{\textwidth}{0.4pt}
\begin{abstract}
\noindent A small excitation gap makes a quantum dot a sensitive probe of a weak electric field, but the charge detector reading it can also excite it. We calculate how much of this enhancement survives continuous readout for two interacting electrons tuned through a confinement-driven crossover in a gate-defined dot. The probe is a few-level system read by a weakly coupled linear detector, with second-order transition rates and information from the mean detector current, for gaps up to a few hundred microelectronvolts, couplings up to 50~\textmu eV and temperatures up to 100~mK. The outcome depends on the detector's energy structure, not only its noise. A quantum point contact measures more strongly at higher bias but, once the bias energy exceeds the gap, also excites the probe; when this back-action dominates the probe's relaxation, the optimised information rate grows only as the inverse of the gap. A sensor quantum dot with a narrow level instead reaches full strength at a bias below the gap, where a non-interacting detector cannot excite it at zero temperature. This energy-filtered readout keeps the inverse-square gap dependence of the quantum Fisher information down to gaps comparable to the detector's level width or ten times the thermal energy. At the tested operating points, higher probe states change the sensitivities by up to 6\% and the exponents by up to 0.06; coherence corrections to the populations and transduced signal are negligible, and any gain from the full detector record is unresolved within 1--2\% numerical uncertainty. Under the same conditions the interacting pair gives about three times the information rate of a single electron with the same gap and detector, for contact and Coulomb-type interactions alike. The different gap dependences and optimal biases of the two detectors can be tested by bias and gate sweeps on charge-sensed double dots.
\end{abstract}

\noindent\rule{\textwidth}{0.4pt}

\section{Introduction}\label{sec:intro}

Criticality-enhanced sensing promises a precision that grows as an energy gap closes~\cite{Zanardi2008,Rams2018,Garbe2020,Montenegro2025,Mihailescu2026tut}. For a weak perturbation the quantum Fisher information (QFI), which bounds the precision of any measurement, acts as a susceptibility, and in nanoelectronic devices it grows as the inverse square of the smallest excitation gap~\cite{Mihailescu2024nano}. Gate-defined quantum dots offer such gaps under direct electrical control. Two interacting electrons driven through the confinement-driven entanglement crossover of~\cite{Abdullah2009,Coe2010} respond strongly to a weak electric field and, as we show below, their QFI exceeds that of a single electron with the same gap.

The literature has also made clear what a critical sensor pays for its gap: preparation time, sensing time, temperature, noise and the restriction to measurements a device can perform~\cite{Rams2018,Gietka2022,Gyhm2026,Mihailescu2026gap,Mihailescu2024qst,Mihailescu2025pra}. One cost has received less attention, although in a solid-state device it cannot be avoided: the detector. A charge detector that reads a gapped probe couples to it through the same transition that carries the signal, and a biased detector can excite that transition. Whether a gap enhancement survives continuous readout is therefore a property of the probe and the detector together.

Each half of this problem is well studied. The continuous monitoring of double-dot charge states by a point contact has been worked out in detail~\cite{Korotkov1999,Korotkov2001,Gurvitz1997,Goan2001}, its linear-detector description is reviewed in~\cite{Clerk2010}, and its inelastic back-action has been derived~\cite{Young2010} and measured~\cite{Gasser2009,Harbusch2010,Granger2012}. The Fisher information of a continuous record can be computed from the stochastic master equation~\cite{Gammelmark2013,Gammelmark2014,Kiilerich2016,Albarelli2018}. The back-action of charge detectors has been measured for point contacts~\cite{Onac2006,Gustavsson2007} and sensor dots~\cite{Horibe2015,Haldar2025}, and studied theoretically for sensor dots acting on relaxation rates~\cite{Schulenborg2014}, on interacting impurities~\cite{Sankar2025} and through many-body correlations in the detector leads~\cite{Grant2026}. In open quantum-optical systems continuous measurement has been proposed as a resource for critical sensing, since counting the quanta emitted by a dissipative critical system can beat a direct measurement of its steady state~\cite{Ilias2022,Ilias2024,Yang2023,Zhang2025}, and the information rate of a sequentially measured probe has been studied in its own right~\cite{OConnor2024}. There the monitored channel is the system's own emission. In a semiconductor device the detector is instead an external, biased conductor that can heat the probe, and monitoring becomes a cost rather than a resource.

\begin{figure}[t]
\centering
\includegraphics[width=\textwidth]{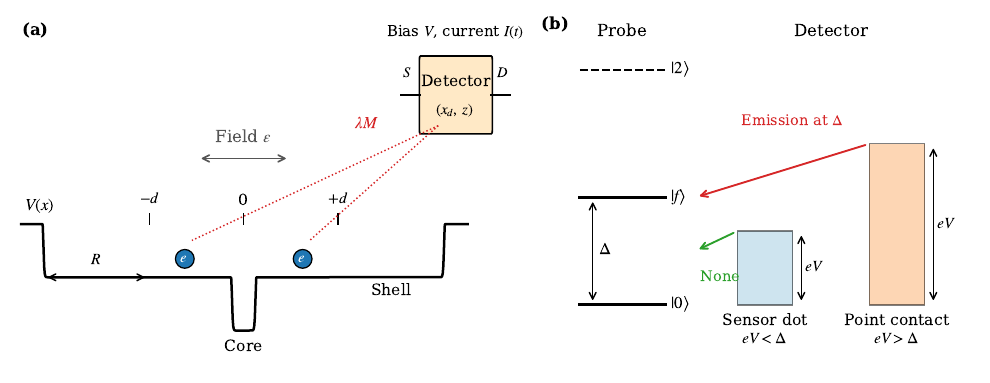}
\caption{(a) The device. Two electrons in the one-dimensional two-centre potential of \Eq{eq:V}, which for $R>d$ has a deep central core and a shallow outer shell, are the probe of a uniform field $\varepsilon$. A charge detector (point contact or sensor dot) placed on one side of the dot, at $(x_d,z)$, is biased by $V$ between its source $S$ and drain $D$; its current $I(t)$ is the record, and its coupling $\lambda M$ to the electrons is also the back-action. (b) The energy picture. The probe has a ground state $|0\rangle$, the field-coupled state $|f\rangle$ at the gap $\Delta$ and higher states ($|2\rangle$). At zero temperature and to second order in the coupling, a non-interacting detector can emit at most the bias energy $eV$ into the probe: a point contact, which needs $eV>\Delta$ to measure quickly, excites the probe, whereas a sensor dot reaches full measurement strength at $eV<\Delta$, where it cannot.}
\label{fig:schematic}
\end{figure}

What has not been done is to put the two halves together for a gapped sensing task. The Fisher information of nanoelectronic probes~\cite{Mihailescu2024nano} was computed for the probe alone, with the detector idealised; the monitoring theory of double dots~\cite{Korotkov1999,Korotkov2001,Gurvitz1997,Goan2001,Clerk2010} treats the readout of a fixed two-level system, not a sensor whose gap is the resource; and the back-action studies~\cite{Young2010,Schulenborg2014,Sankar2025,Grant2026} ask how a detector perturbs a dot, not how much field information the detector can deliver as the gap closes. This paper closes that gap. It answers the question above for the two-electron crossover and for the two charge detectors used in practice, a quantum point contact and a sensor quantum dot~\cite{Field1993,Cassidy2007,Reilly2007,Barthel2010} (\Fig{fig:schematic}), and it does so with the detector's own energy structure, bias and temperature in the calculation. The outcome is decided by the energy structure of the detector, not only by its noise. (i) Parity selects the readout: a detector placed symmetrically about the dot centre has no linear response at the symmetric operating point, so the signal needs a detector on one side, which couples to the field-carrying transition with an amplitude equivalent to about half an electron. (ii) Through such a detector the signal scales as the square of the transition dipole divided by the gap, so the information scales as the fourth power of the dipole, and the advantage of the interacting pair over a single electron with the same gap rises from 1.7 in the QFI to about 3. (iii) A detector whose measurement strength grows with its bias, and whose back-action spectrum has the bias as its only energy scale, halves the gap exponent of the information rate whenever its back-action dominates the intrinsic relaxation of the probe. (iv) A detector whose measurement strength saturates at a bias below the gap, such as a sensor dot, can measure at full strength without emitting at the gap; such an energy-filtered detector keeps the inverse-square law for any weak coupling, down to a floor set by its level width and by the electron temperature. This needs neither a weaker coupling nor a quieter detector, only a bias below the gap; it rests on the absence of emission above the bias in the non-interacting detectors considered, to second order in the coupling. (v) The two laws give a direct experimental signature in the slope of the sensitivity against the gap. What is new, relative to the current-based sensing analysis of~\cite{Mihailescu2024nano}, is that the optimised gap scaling depends on the energy structure of a separate detector; relative to the monitoring literature, it is that the quantity optimised is field information rather than readout fidelity; and relative to the back-action literature, it is the comparison of the two detectors on equal terms over a range of gaps, temperatures and couplings, with the range over which the calculation is controlled stated explicitly.

Sections~\ref{sec:probe} and~\ref{sec:parity} introduce the probe and its coupling to a detector, sections~\ref{sec:rate}--\ref{sec:scaling} develop the detector theory and test the two scaling laws, \Sec{sec:pair} compares the pair with a single electron, and \Sec{sec:discussion} gives the predictions, the dependence of the results on the model and its parameters, and the limits of the model. The checks of the two-level rate description, namely the effect of the higher states of the probe, of the coherences neglected in the rate equations, of the information left in the full detector record and of Coulomb blockade in the sensor dot, are collected in \App{app:checks}.

\section{The probe}\label{sec:probe}

We use the model of~\cite{Abdullah2009,Coe2010}: two electrons in one dimension in the two-centre power-exponential potential
\begin{equation}
V(x)=-V_0\left\{\exp\!\left[-\left(\frac{|x+d|}{R}\right)^{p}\right]+\exp\!\left[-\left(\frac{|x-d|}{R}\right)^{p}\right]\right\},
\label{eq:V}
\end{equation}
with $V_0=10$, $d=8$ and $p=200$ in effective atomic units. These are the atomic units of an electron of effective mass $m^{*}$ in a medium of relative permittivity $\epsilon_r$, $\hbar=m^{*}=e^{2}/4\pi\epsilon_0\epsilon_r=1$: the unit of length is the effective Bohr radius $a^{*}=a_0\epsilon_r/(m^{*}/m_e)$, the unit of energy the effective Hartree $\Ha=\mathrm{Ha}\,(m^{*}/m_e)/\epsilon_r^{2}$, the unit of electric field $\Ha/ea^{*}$ and the unit of time $\hbar/\Ha$. For GaAs, with $m^{*}=0.067\,m_e$ and $\epsilon_r=12.4$, $a^{*}=9.79$~nm, $\Ha=11.86$~meV, one unit of field is $1.21\times10^{6}$~V\,m$^{-1}$ and one unit of time is 55.5~fs, so that a rate of 1~meV$/\hbar$ is $1.52\times10^{12}$~s$^{-1}$. The well centres are then $2d=157$~nm apart. All lengths, energies and matrix elements below are quoted in these units unless a physical unit is given, and the detector calculations of \App{app:detectors} are done in meV after this conversion.
The Hamiltonian is
\[
H=\sum_{i=1,2}\left[-\tfrac{1}{2}\partial_{x_i}^{2}+V(x_i)+\varepsilon x_i\right]+U(x_1-x_2),
\]
where $\varepsilon$ is the electric field and $U$ is a contact interaction, $U=g\,\delta(x_1-x_2)$ with $g=1$ unless stated, or a Coulomb-type interaction (\Sec{sec:robust}). We work in the spin-singlet sector, which contains the ground state for any symmetric interaction in one dimension~\cite{Lieb1962}. At the working point ($R=8.035$) the lowest triplet lies 90~\textmu eV above the ground state, so its equilibrium weight at 100~mK is $10^{-4}$; along the interaction scan of \Sec{sec:scaling} this gap falls to 10~\textmu eV, and there we assume a protocol that starts in the singlet and conserves spin.

The range $R$ deforms the structure from a double well ($R<d$) into a single well with a deep central core and a shallow outer shell ($R>d$). At a sharp value of $R$ the ground state switches from a state with one electron in each well to a state in which both electrons occupy the core. We call this avoided crossing between two even ground-state branches the crossover. With only two particles, its sensing figures saturate rather than diverge as the walls are made steeper (the peak QFI, the gap and the dipole no longer change for $p\gtrsim100$), so we do not call it a phase transition.

\begin{figure}[t]
\centering
\includegraphics{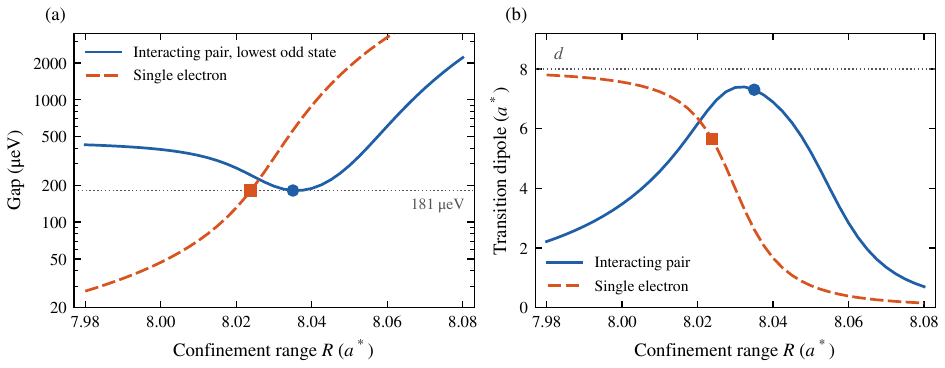}
\caption{The crossover ($p=200$, $d=8$, contact interaction). (a) Gap of the lowest odd state of the interacting pair, identified by parity at each $R$ and followed from one $R$ to the next, and of a single electron in the same confinement, as functions of the confinement range $R$. The dotted line marks the matched gap of 181~\textmu eV; the circle and the square mark the working points. (b) Transition dipole to the same state. Between $R=8.01$ and 8.06 this single state carries more than 92\% of the QFI for the field (99.95\% at the working point). The single electron opens its gap by delocalising and loses its dipole, whereas the pair opens the same gap through its even anticrossing and keeps it.}
\label{fig:crossover}
\end{figure}

The eigenstates are obtained by configuration interaction: single-particle orbitals computed on a real-space grid ($|x|\le26$, spacing 0.01) span the singlet space of the lowest $n_{\mathrm{orb}}$ orbitals, and the interaction is evaluated by quadrature ($n_{\mathrm{orb}}=50$ unless stated; \App{app:ci}). Its validation tests are described in \App{app:ci}. A uniform field couples through the dipole $X=x_1+x_2$, which is odd under inversion, so only odd states contribute to the field response. At the working point $R^{*}=8.035$ a single odd state $|f\rangle$ carries 99.95\% of it, so that $F_Q\simeq4\Xf^{2}/\Delta^{2}$, with $\Delta=E_f-E_0=1.528\times10^{-2}\,\Ha=181$~\textmu eV and $\Xf=7.31\,a^{*}$. The states are identified as follows. At each $R$ the solver returns the lowest ten singlets and the dipole and detector matrices between them; the field-coupled state $|f\rangle$ is the excited singlet with the largest $|X_{0n}|$ among the lowest eight, and parity is read off from the orbital content of each state. In \Fig{fig:crossover} the lowest odd state is followed by parity from one $R$ to the next; between $R=8.00$ and 8.07 it coincides with $|f\rangle$, and the ground state is followed as the lowest eigenvalue, which is continuous through the even anticrossing. The next singlets at the working point lie at 0.55~meV (even) and 0.76~meV (even) above the ground state; their role in the readout is examined in \App{app:checks}.

\FIG{fig:crossover} compares the pair with a single electron in the same confinement. As the figure shows, the even anticrossing reshapes the ground state and raises its dipole coupling to the odd state from about 2 to 7.3~$a^{*}$, while the odd gap that governs the field response passes through a shallow minimum and never closes. A single electron can open its gap only by delocalising, which costs dipole: tuned to the same gap ($R_1=8.0238$) it has $X_{01}=5.65\,a^{*}$, so the pair holds 1.67 times its QFI. We use this matched single electron as the reference throughout, since at unequal gaps the comparison would be dominated by whichever probe has the smaller gap: the QFI of a two-level probe grows as $\Delta^{-2}$ whatever its dipole.

\section{Parity selects the readout}\label{sec:parity}

A charge detector at position $(x_d,z)$ couples to the electrons through the one-body operator $M=\sum_i f(x_i)$. We take the unscreened Coulomb kernel of a point charge, $f(x)=[(x-x_d)^2+z^2]^{-1/2}$, and for comparison an ideal which-side kernel $f=\theta(x)$ and two symmetric controls. Matrix elements are quoted in units of the one-electron transfer contrast $f(d)-f(-d)$, the unit in which charge sensitivities are specified~\cite{Cassidy2007,Barthel2010}.

\begin{table}[t]
\caption{Detector matrix elements of $M=\sum_if(x_i)$ at the working point ($R=8.035$, $g=1$, 50 orbitals). The detector position $(x_d,z)$ is in units of $a^{*}$; $c=f(d)-f(-d)$ is the one-electron transfer contrast of the kernel in units of $1/a^{*}$ (1 for the which-side kernel), and the matrix elements $M_{00}$, $M_{ff}$, $\mu=M_{0f}$, $\delta=M_{ff}-M_{00}$ and the elements to the next singlet $|2\rangle$ are quoted in units of $c$. For the symmetric kernels the contrast vanishes, so their unnormalised $M_{0f}$ is shown as a symmetry check; it is zero to machine precision.}
\label{tab:kernels}
\centering
\footnotesize
\setlength{\tabcolsep}{4pt}
\begin{tabular}{lccccccccc}
\toprule
Kernel & $c$ & $M_{00}$ & $M_{ff}$ & $\mu$ & $\delta$ & $\delta/2\mu$ & $M_{02}$ & $M_{f2}$ & $M_{22}$\\
\midrule
Uniform, $f=1$ & 0 & 2 & 2 & $<10^{-15}$ & 0 & -- & 0 & 0 & 2\\
Symmetric, $f=(x^2+z^2)^{-1/2}$, $z=6$ & 0 & -- & -- & $<10^{-12}$ & -- & -- & -- & -- & --\\
Which-side, $f=\theta(x)$ & 1 & 1.001 & 1.001 & 0.433 & 0.000 & 0.00 & 0.000 & $-0.190$ & 1.001\\
Coulomb, $(12,4)$ & 0.1277 & 1.477 & 1.588 & 0.461 & 0.111 & 0.12 & $-0.173$ & $-0.366$ & 1.540\\
Coulomb, $(16,6)$ & 0.0596 & 2.161 & 2.265 & 0.482 & 0.104 & 0.11 & $-0.157$ & $-0.396$ & 2.250\\
Coulomb, $(24,8)$ & 0.0256 & 3.232 & 3.307 & 0.474 & 0.076 & 0.08 & $-0.114$ & $-0.378$ & 3.300\\
Coulomb, $(40,12)$ & 0.0090 & 5.380 & 5.425 & 0.463 & 0.045 & 0.05 & $-0.068$ & $-0.355$ & 5.418\\
\bottomrule
\end{tabular}
\end{table}

\TAB{tab:kernels} gives the first result. Both symmetric kernels give $M_{0f}=0$ to machine precision at every $R$, because $|0\rangle$ and $|f\rangle$ have opposite parity while a symmetric $M$ commutes with parity. A symmetric detector, including the quantum capacitance seen by a gate placed symmetrically about the dot centre, is therefore blind to a weak field at the symmetric point. Every one-sided kernel gives $M_{0f}=0.43$--0.48, close to $\Xf/2d$, and the single state $|f\rangle$ carries 99.7\% of the static response $\partial\langle M\rangle/\partial\varepsilon$. Unless stated otherwise we use the Coulomb kernel with the detector at $(x_d,z)=(12,4)$; with the gap, the dipoles and the elements of \Tab{tab:kernels}, the two-level detector calculations of this paper can be repeated from the formulas of \App{app:detectors} without the solver; the energies, dipole and detector matrices of the lowest eight singlets of both probes, needed for \App{app:multilevel}, are in the archive.

In the $\{0,f\}$ subspace we write $M=m_0+\tfrac{\delta}{2}\sigma_z+\mu\sigma_x$, with $\mu=M_{0f}$ and $\delta=M_{ff}-M_{00}$. The diagonal contrast $\delta$ distinguishes the two states without changing them, whereas the off-diagonal element $\mu$ drives transitions between them. For the one-sided Coulomb kernels $\delta/2\mu$ is small (0.05--0.12, and zero for the which-side kernel): the detector couples mainly through the part that drives transitions, so detector-induced excitation is intrinsic to this readout and not the mark of a poorly placed detector. The mean detector charge also tilts the probe statically (\Sec{sec:rate}); as in any charge-sensed device, we assume that this tilt is compensated when the symmetric point is calibrated with the gates, so that the probe stays at its parity-symmetric operating point.

\section{Transduction and the information rate}\label{sec:rate}

In the $\{0,f\}$ subspace the probe Hamiltonian is $H=\tfrac{\Delta}{2}\sigma_z+\varepsilon\Xf\sigma_x$. To first order in the field the dressed eigenstates $|0'\rangle$ and $|f'\rangle$ give
\begin{equation}
M_{0'0'}=M_{00}-a\varepsilon,\qquad M_{f'f'}=M_{ff}+a\varepsilon,\qquad a\equiv\frac{2\mu\Xf}{\Delta},
\label{eq:transduction}
\end{equation}
so the field shifts the mean detector signal in opposite directions in the two states. Since $\mu\simeq\Xf/2d$, this transduced signal scales as $\Xf^{2}/\Delta$: the transition dipole enters twice, once through the coupling to the field and once through the detector's view of the transition. This is the structure of a dispersive shift in circuit quantum electrodynamics, here with the two couplings to different fields.

We treat each detector as a linear detector~\cite{Clerk2010} coupled through $H_{\mathrm{int}}=\lambda MQ$, where $Q$ is the detector's charge variable and $\lambda$ is the energy by which one electron transferred across the probe shifts the detector's scattering energy; the mean term $\lambda\langle Q\rangle M$ is the static tilt of \Sec{sec:parity}. With the back-action spectrum $S(E)=\hbar^{-1}\int\mathrm{d}t\,\mathrm{e}^{\mathrm{i}Et/\hbar}\langle\delta Q(t)\,\delta Q(0)\rangle$, where $\delta Q=Q-\langle Q\rangle$ and $E<0$ corresponds to emission by the detector, the transition rates to second order in the coupling are
\begin{equation}
\Gup=\frac{\lambda^2\mu^2}{\hbar}S(-\Delta)+\Gamma_0 n_B,\qquad
\Gdn=\frac{\lambda^2\mu^2}{\hbar}S(+\Delta)+\Gamma_0(1+n_B),\qquad
\Gamma_\phi=\frac{\lambda^2\delta^2}{2\hbar}S(0).
\label{eq:rates}
\end{equation}
Here $S(E)$ is the non-symmetrised spectrum of the detector's charge fluctuations: $S(E>0)$ is the weight with which the detector absorbs the energy $E$ from the probe and $S(E<0)$ the weight with which it emits $|E|$ into it, so that $\Gup$ needs $S(-\Delta)$ and $\Gdn$ needs $S(+\Delta)$, and $S(0)$ sets the dephasing. $\Gamma_0=1/T_1$ is an intrinsic charge relaxation rate, with $T_1=16$~ns, the value measured in a GaAs double dot~\cite{Petta2004}, taken as a representative scale, and $n_B=[\mathrm{e}^{\Delta/k_BT}-1]^{-1}$ is the Bose occupation at the probe temperature, which in the numerical examples is taken equal to the electron temperature of the detector leads. The detector output $I$ responds to $M$ with gain $\lambda\,\partial I/\partial\epsilon_0$, where $\epsilon_0$ is the detector's scattering energy, and has zero-frequency noise $S_{II}=\int\mathrm{d}t\,\langle\delta I(t)\,\delta I(0)\rangle$, the two-sided covariance intensity, which equals the symmetrised spectral density of~\cite{Clerk2010} at zero frequency and is half the one-sided spectral density of engineering convention. The imprecision referred to the probe is $S_q=S_{II}/(\lambda\,\partial I/\partial\epsilon_0)^{2}$, the white noise that must be added to $M$ to reproduce the output noise; the measurement rate is $\Gmeas=\delta^2/8S_q$, and $\eta=\Gmeas/\Gamma_\phi\le1$ is the quantum efficiency~\cite{Clerk2010} (\App{app:detectors}).

Because $\Delta/\hbar\simeq2.7\times10^{11}$~s$^{-1}$ is large compared with all these rates, coherences between the two states average out, and the monitored probe jumps at random times between $|0'\rangle$ and $|f'\rangle$, a telegraph process with stationary excited population $p_f=\Gup/(\Gup+\Gdn)$. The long-time Fisher information of the time-averaged detector current per unit time, which we call the mean-current information rate, is
\begin{equation}
\Fm=\frac{(\partial_\varepsilon\bar M)^2}{S_q+S_{\mathrm{tel}}},\qquad
\partial_\varepsilon\bar M=-a(1-2p_f)+\delta\,\partial_\varepsilon p_f,\qquad
S_{\mathrm{tel}}=\frac{2\delta^2p_f(1-p_f)}{\Gup+\Gdn}.
\label{eq:Fmean}
\end{equation}
The factor $1-2p_f$ appears because the two dressed states shift the signal in opposite directions, and the field enters the populations through $M_{0'f'}=\mu-\varepsilon\Xf\delta/\Delta$. $S_{\mathrm{tel}}$ is the telegraph noise, the zero-frequency spectrum of the random jumps of the signal between its two levels $\pm\delta/2$ (two-sided, in the same convention as $S_q$), which adds to the imprecision because a time average of the current cannot tell a jump from noise. By the data-processing inequality, $\Fm$ is a lower bound on the information rate of the full record. For a one-sided point contact the jumps are hard to resolve: the population measurement ($\propto\delta^2$) and the switching it would have to resolve ($\propto\mu^2$) are driven by the same detector noise, so the ratio of measurement rate to switching rate obeys $\Gmeas/(\Gup+\Gdn)\le(\eta/4)(\delta/\mu)^2\simeq0.015\,\eta$ and individual switching events are not resolved. We have quantified what the full record adds by computing its Fisher information numerically for the telegraph model (\App{app:fullrecord}): at the optimised operating points of the point contact no excess information in the full record is resolved within the 1--2\% numerical uncertainty of the calculation. For a sensor dot biased below the gap $\delta/\mu$ is the same but the switching rates are far smaller, and at the working gap and 10~mK the ratio of measurement rate to switching rate reaches 0.05--0.6 for the widths studied and $\lambda=20$--50~\textmu eV; since the probe is then almost never excited ($p_f<10^{-6}$), there are no jumps to resolve, and a bound on the information of a record whose state path is known exactly shows that the full record can exceed the mean current by about $10^{-6}$ or less at the working point (\App{app:fullrecord} gives the bound along the scan). The two-level telegraph description is itself a model assumption; its checks are given in \App{app:checks} and its remaining limits in \Sec{sec:limits}.

\section{The two detectors}\label{sec:detectors}

Both detectors are single-channel conductors whose transmission is shifted by $\lambda M$, and we describe both with the same coupling $\lambda$ so that they can be compared on equal terms (\App{app:detectors}).

\textit{Point contact.} The point contact has the saddle-point transmission $\Tr(E)=[1+\mathrm{e}^{-2\pi(E-E_0)/\hbar\omega_x}]^{-1}$ with $\hbar\omega_x=1$~meV, set to transmission $\tau=1/2$; its current, noise and gain are integrated over the bias window, so the flattening of the transmission at a few hundred microvolts is included. For its back-action we take the spectrum of Aguado and Kouwenhoven~\cite{Aguado2000}, whose emission side is cut off at $\hbar\omega=eV$~\cite{Gustavsson2007}, normalised by requiring the detector to be quantum limited ($\eta=1$) at small bias and zero temperature. This is an effective model (\Sec{sec:limits}): the scaling law of \Sec{sec:scaling} uses only the homogeneity of the spectrum in the bias, whereas the absolute numbers depend on the model. The essential property is that both the gain and the noise of a point contact grow with bias, so $\Gmeas$ grows with $eV$ until the transmission flattens: a point contact cannot measure quickly without a large bias.

\textit{Sensor dot.} We model the sensor dot as a spinless, non-interacting resonant level of width $\Gamma$ between two symmetrically biased leads, with its level at energy $\epsilon_0$, so that its current, noise, gain and charge-noise spectrum are exact to all orders in $\Gamma$. Such a detector reaches the quantum limit only with its level on the flank of the resonance, since at resonance the gain vanishes while the dephasing does not~\cite{Khym2006,Clerk2003,Pilgram2002}; our optimiser places the level on the flank, and $\eta\le1$ everywhere, with $\eta\to1$ at small bias and zero temperature. The essential property here is that the current saturates once the bias exceeds $\Gamma$. The gain is then set by the level width rather than by the bias, so $\Gmeas\propto\lambda^{2}/\Gamma$ independently of $eV$ for $eV\gtrsim\Gamma$, provided the level sits within about $\Gamma$ of a lead chemical potential, which is what the optimisation does and what an experiment does with the plunger gate.

What the two detectors share is energy conservation. In the non-interacting, dc-biased detectors considered here a transmitted electron can supply at most $eV$, so at zero temperature the emission spectrum at the gap vanishes for $eV<\Delta$, $S(-\Delta)=0$, and detector-induced excitation vanishes to second order in the coupling. This threshold has been observed with double-dot photon detectors~\cite{Gustavsson2007,Onac2006} and derived for a point contact reading a double dot~\cite{Ouyang2010}. It is a property of these uncoupled detector spectra: correlations inside an interacting detector or its environment can create spectral weight above the bias, as over-bias emission in tunnel junctions shows~\cite{Xu2014}. At finite temperature the thermal tails of the lead distributions restore a small emission, of order $\mathrm{e}^{-(\Delta-eV)/k_BT}$. The two detectors differ in what a bias below the gap costs them.

\begin{figure}[t]
\centering
\includegraphics{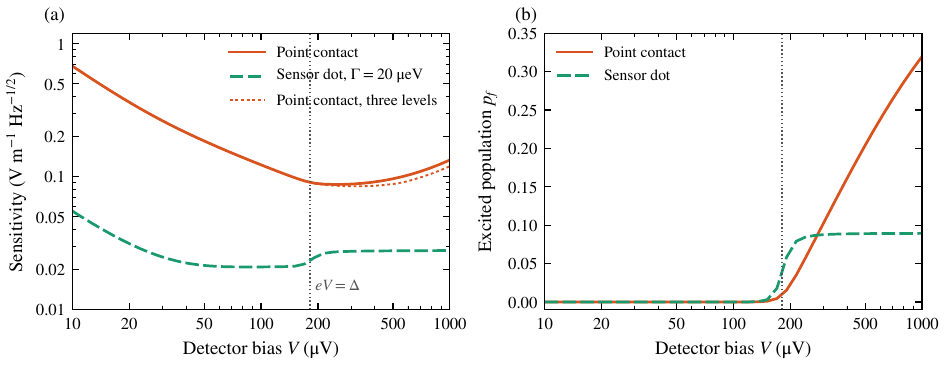}
\caption{Bias sweep at the working gap ($\Delta=181$~\textmu eV), 100~mK and $\lambda=50$~\textmu eV. (a) Field sensitivity $\Fm^{-1/2}$ for the point contact and for a sensor dot of width $\Gamma=20$~\textmu eV, with the level position optimised at each bias; the dotted orange line is the point contact with the next singlet of the probe included in the rate equations (\App{app:multilevel}). (b) Excited population $p_f$ of the probe in the two-level model. The vertical dotted line marks the emission threshold $eV=\Delta$.}
\label{fig:bias}
\end{figure}

To see what this means at the working gap, \Fig{fig:bias} shows the sensitivity and the excited population of the probe as functions of the detector bias. As the figure shows, the point contact improves with bias up to the emission threshold and then pays for further bias with excitation: its best operating point lies above the threshold, at 250~\textmu V, where $p_f\simeq0.07$ and the factor $(1-2p_f)^2$ starts to erode what the bias buys. The sensor dot reaches its best sensitivity at a bias of a few tens of microvolts, well below the gap, with an excited population of about $2\times10^{-7}$ (the optimum is flat, and $p_f$ varies by more than an order of magnitude across it). At equal coupling its sensitivity is four times better than that of the point contact (0.021 against 0.088~V\,m$^{-1}$\,Hz$^{-1/2}$). The dotted line in \Fig{fig:bias}(a) shows the point contact when the next singlet of the probe is included in the rate equations (\App{app:multilevel}): above about 0.4~mV the bias can also excite it, the optimum moves from 250 to 320~\textmu V and the best sensitivity changes by 3\%; for the sensor dot the higher states are never populated. Suppressing excitation does not remove back-action: the sensor dot still dephases the probe, at $\Gamma_\phi\simeq4\times10^{8}$~s$^{-1}$, and relaxes it. Within our model neither affects a protocol that keeps the probe in its ground state. The detector also renormalises the probe statically, through the second-order (Lamb) shift of its levels; this shift is evaluated in \App{app:coherences} and lowers the gap by 1~\textmu eV at $\lambda=20$~\textmu eV and by 6~\textmu eV at $\lambda=50$~\textmu eV, a change of at most 3.5\% that is absorbed in the gate calibration of the working point. The residual excitation at 100~mK comes from the thermal tails of the detector leads (a rate of $1.3\times10^{2}$~s$^{-1}$), not from the intrinsic bath ($0.05$~s$^{-1}$).

\section{The gap scaling}\label{sec:scaling}

\subsection{A detector with a single energy scale halves the exponent}\label{sec:single}

Combining \Eq{eq:Fmean} with the efficiency relation $S_q=\delta^2/(8\eta\Gamma_\phi)$ gives, when the imprecision dominates the telegraph noise and the detector dominates the intrinsic relaxation,
\begin{equation}
\Fm=\frac{4\eta\lambda^2a^2}{\hbar}\,S(0;V)\,\Asym(V)^{2},\qquad
\Asym\equiv\frac{S(+\Delta)-S(-\Delta)}{S(+\Delta)+S(-\Delta)},
\label{eq:Fsingle}
\end{equation}
that is, the measurement strength times the squared asymmetry of the detector noise at the gap of the probe. Now suppose that (i) the detector dominates the relaxation, (ii) $\lambda$ and $\eta$ are fixed, (iii) the two-level telegraph description holds, and (iv) at zero temperature the back-action spectrum has the bias as its only energy scale, $S(E;V)=eV\,s(E/eV)$. Writing $eV=v\Delta$, both $S(\pm\Delta)$ become $v\Delta\,s(\pm1/v)$, and
\begin{equation}
\max_{V}\Fm=C\,a^{2}\Delta,\qquad C=\frac{4\eta\lambda^{2}}{\hbar}\,\max_{v}\left[v\,s(0)\,\Asym(v)^{2}\right].
\label{eq:Cmax}
\end{equation}
The trade-off between buying measurement strength with bias and the emission that the same bias enables has no scale of its own, so the only scale left is the gap. With $a\propto\Xf^{2}/\Delta$,
\begin{equation}
\Fm^{\max}\propto\frac{\Xf^{4}}{\Delta}\qquad\textrm{against}\qquad F_Q\propto\frac{\Xf^{2}}{\Delta^{2}}.
\label{eq:law1}
\end{equation}
Under assumptions (i)--(iv), detector back-action therefore halves the gap exponent of the mean-current information rate. For the point-contact spectrum at half transmission and with a flat transmission the optimum lies at the threshold itself, $eV^{*}=\Delta$, and $C$ contains the pure number $1/4$; any other homogeneous spectrum for which $v\,s(0)\Asym(v)^2$ has a finite positive maximum gives the same exponent with a different prefactor and optimum. Under the same assumptions, and for an explicit observation model of the record, the inverse-gap law holds for the information rate of the whole record and not only for its mean (\App{app:detectors}). The degree of homogeneity matters: a measurement strength $S(0;V)\propto(eV)^{r}$ would give the exponent $r-2$ instead of $-1$. Within the electrical model the law is broken by condition (i): a detector too weak to compete with $T_1$ keeps $p_f$ small at any bias, is limited by the largest bias it can take rather than by its back-action, and shows an exponent between 1 and 2 with a much smaller absolute information.

\subsection{An energy-filtered detector keeps the exponent}\label{sec:filtered}

For the sensor dot condition (iv) fails in the way that matters. Its measurement strength saturates at $eV\sim\Gamma$, so the optimisation of \Eq{eq:Fsingle} no longer trades strength against emission. For $\Gamma\ll\Delta$ the detector can be biased at $\Gamma\lesssim eV<\Delta$, where $S(-\Delta)=0$, $\Asym=1$ and $S(0)$ has already reached its maximum. We call such a detector energy filtered: it measures at full strength while it cannot emit at the gap. Then
\begin{equation}
\Fm^{\max}\simeq\frac{4\eta\lambda^2a^2}{\hbar}\,S(0)\propto\frac{\lambda^{2}}{\Gamma}\,\frac{\Xf^{4}}{\Delta^{2}},
\label{eq:law2}
\end{equation}
and the closed-system exponent returns for any coupling $\lambda$ within the weak-coupling regime of \Eq{eq:rates}, because the coupling multiplies both the measurement strength and an excitation rate that is zero at this order. The floor of this regime is set by the detector's own energy scales. When $\Delta\lesssim\Gamma$ the sensor dot needs a bias above the gap to reach its gain and behaves like a detector with a single energy scale. When $k_BT$ is not small compared with $\Delta$, the thermal tails of the lead distributions restore emission below the threshold, with a rate that depends exponentially on $(\Delta-eV)/k_BT$ and that already exceeds the intrinsic excitation by three orders of magnitude at the working point and 100~mK (\Sec{sec:detectors}).

\subsection{Numerical test}\label{sec:test}

\begin{figure}[t]
\centering
\includegraphics{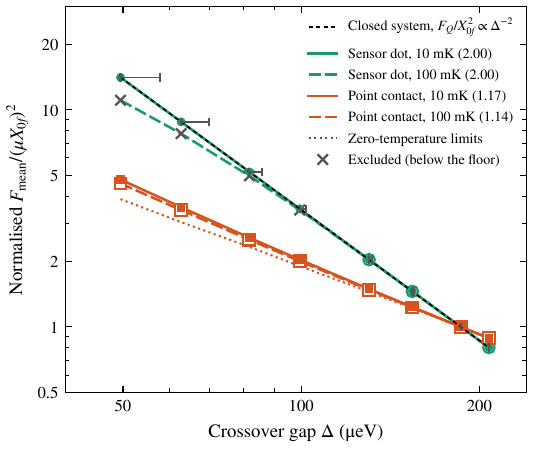}
\caption{Mean-current information rate against the crossover gap, tuned by the interaction strength ($g=0.7$--8) and taken at the basis-extrapolated gap and matrix elements of each working point (\Tab{tab:convergence}; the horizontal bars on the two smallest gaps show the difference between the linear and quadratic extrapolations), divided by $(\mu\Xf)^2$ and normalised to its value at $g=1$ ($\Delta=186$~\textmu eV). Sensor dot: $\Gamma=\lambda=20$~\textmu eV; point contact: $\lambda=50$~\textmu eV; $T_1=16$~ns; bias and level re-optimised at each gap. Filled markers: 10~mK; open markers: 100~mK; crosses: points below the sensor-dot floor, excluded from the fits. Numbers in brackets are the fitted exponents. The dashed black line is the closed-system QFI, $F_Q/\Xf^2\propto\Delta^{-2}$, and the dotted lines are the zero-temperature limits dominated by the detector.}
\label{fig:scaling}
\end{figure}

We test both laws in the interacting model by tuning the crossover gap through the interaction strength $g$, relocating the working point $R^{*}(g)$ at each $g$ to the maximum of $4\Xf^2/\Delta^2$. This is a numerical device for varying the gap, not a proposed control. The working points are located with 50 orbitals, and the gap, the dipole and the detector matrix elements at each of them are then followed from 50 to 150 orbitals and extrapolated in $1/n_{\mathrm{orb}}$ (\App{app:ci}, \Tab{tab:convergence}); these basis-extrapolated estimates, whose uncertainty reaches 16\% in the gap at the strongest interaction, are the ones used here and in \Fig{fig:scaling}, and the exponents are also given for the other bases below. Between $g=0.7$ and $g=8$ the gap falls from 208 to 50~\textmu eV, a factor of 4.2, while $\Xf$ grows from 7.2 to 7.8~$a^{*}$ and $\delta/2\mu$ falls from 0.12 to 0.05. At each point we optimise the bias, and for the sensor dot also the level position, at fixed $\lambda$, and we divide the information rate by $(\mu\Xf)^2$ to isolate its dependence on the gap.

Two criteria decide which points enter the fits. The telegraph description requires the total switching rate to be small compared with the gap, so we keep only points with $\hbar(\Gup+\Gdn)/\Delta<0.1$; a small excited population does not guarantee this, because absorption and relaxation remain large when emission is suppressed. With the basis-extrapolated gaps this criterion is met at every point: the ratio stays below 0.02 for the sensor dot at $\lambda=20$~\textmu eV and below 0.1 at $\lambda=50$~\textmu eV (0.096 at the smallest gap), and below 0.003 for the point contact. In addition, the sensor-dot law requires the gap to lie above the floor of \Sec{sec:floor}. Dephasing is small throughout ($\hbar\Gamma_\phi/\Delta\le1.6\times10^{-3}$), and for the sensor dot the switching rate is also small compared with the detector linewidth, $\hbar(\Gup+\Gdn)/\Gamma\le0.04$ at $\lambda=20$~\textmu eV. These diagnostics support the secular, Markovian description, and \App{app:coherences} tests it directly: the non-secular Bloch--Redfield equation with the same spectra changes the stationary populations and the transduced signal by less than $10^{-7}$ at five of the operating points of \Tab{tab:matched} (listed in \Sec{sec:limits}). What the diagnostics do not control is the validity of the second-order rates themselves at the strongest couplings. Their expansion parameter is the transverse coupling $\lambda\mu/\Delta$, which is 0.13 at the working point for $\lambda=50$~\textmu eV and grows to 0.38 and 0.49 at the two strongest-interaction points (0.15 and 0.20 at $\lambda=20$~\textmu eV); the natural estimate of their first corrections, heuristic rather than computed, is of relative order $(\lambda\mu/\Delta)^2$, that is 2\% at the working point but 15--24\% at those two points for $\lambda=50$~\textmu eV. This is why the sensor dot, whose exponent is the more sensitive to its rates, is run at $\lambda=20$~\textmu eV, for which $(\lambda\mu/\Delta)^2\le0.04$ over the whole scan, why the point contact is run at $\lambda=50$~\textmu eV only because its back-action must dominate $T_1$ for the law to apply, and why the exponents are also quoted with the two strongest-interaction points removed.

The exponents are obtained as follows. For each detector, coupling and temperature the information rate is optimised over the bias (and the level position) at each of the eight gaps, divided by $(\mu\Xf)^2$, and a straight line is fitted by least squares to $\ln[\Fm/(\mu\Xf)^2]$ against $\ln\Delta$ over the points that pass the two criteria; the exponent is minus the slope, and its uncertainty is the standard error of the slope. The gaps span 50--208~\textmu eV, a factor of 4.2, so a slope of 1 corresponds to a factor of 4.2 in the normalised information and a slope of 2 to a factor of 18. A local exponent on a finer grid of gaps, which shows where each law holds, is given in \Sec{sec:floor}.

\FIG{fig:scaling} shows the result. In the zero-temperature limit dominated by the detector the fitted exponents over all eight points are $1.023\pm0.003$ for the point contact and $2.000\pm0.001$ for the sensor dot, as \Eqs{eq:law1}{eq:law2} predict. With intrinsic relaxation included, the point contact gives $1.17\pm0.01$ at 10~mK and $1.14\pm0.01$ at 100~mK over all eight gaps, and $1.14$--$1.15\pm0.01$ and $1.12\pm0.01$ over the six largest, where its back-action dominates most clearly. Its optimum bias lies above the threshold and moves with the gap ($eV^{*}=1.2\Delta$--$2.7\Delta$ at 10~mK). The sensor dot gives $2.000\pm0.001$ at 10~mK over all eight gaps, which all lie above its floor of 29~\textmu eV, and $1.999\pm0.001$ at 100~mK over the four gaps above the floor of 101~\textmu eV, $g=0.7$--2, that is 130--208~\textmu eV, a factor of only 1.6 in the gap ($1.85\pm0.04$ over all eight, the four below the floor pulling it down); at $\lambda=50$~\textmu eV the same fits give 2.00 and 2.00. Its optimum bias does not follow the gap: at 10~mK the optimiser places it at about 24~\textmu V, of the order of $\Gamma$, at every gap, within a flat optimum. At $\lambda=20$~\textmu eV the point contact is limited by its bias rather than by its back-action, and its exponent moves to about 1.6, at the price of a factor of two to five in absolute information. With the next singlet of the probe included in the rate equations the point-contact exponents at $\lambda=50$~\textmu eV change by $-0.01$ and $-0.02$ and with the next two by $-0.03$ and $-0.04$, and the sensor-dot exponents do not change (\App{app:multilevel}). For the other couplings and widths, the points used in each fit are listed in the archive.

The basis convergence of the probe matters here and nowhere else in the paper, because the strong-interaction points converge slowly (\App{app:ci}). With the 50-orbital values of the working points the same fits give 1.22 and 1.15 for the point contact and 2.00 and 2.00 for the sensor dot; with the 150-orbital values 1.19 and 1.15 and 2.00 and 2.00; and with the quadratic instead of the linear extrapolation 1.17 and 1.14 and 2.00 and 1.99, the last because the $g=3$ gap then lies at 102~\textmu eV, just above the 100~mK floor, and enters the fit. The basis error moves a point along the sensor-dot law rather than off it, because the information rate of the sensor dot depends on the probe only through $\Delta$ and $\mu\Xf$ above the floor, and it changes the membership of points near the floor; for the point contact the unconverged, too-small gaps of the two strongest-interaction points steepen the fit at 10~mK by 0.05, and the two extrapolation forms differ by 0.006. Dropping those two points leaves the sensor-dot exponents at 2.00 and moves the point-contact exponents to 1.15 and 1.12 at 10 and 100~mK with every basis, so the contrast between the detectors does not depend on the fit range or on the basis.

\subsection{The floor}\label{sec:floor}

\begin{figure}[t]
\centering
\includegraphics{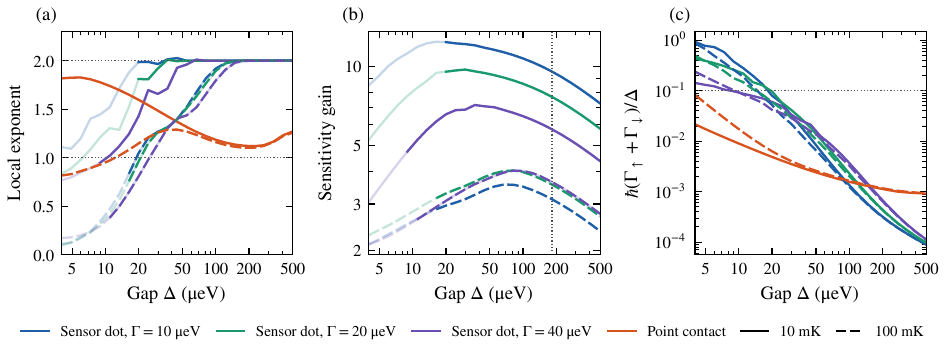}
\caption{Range of validity of the two laws, at the matrix elements $\mu$, $\delta$ and $\Xf$ of the working point, on a 25-point scan of the gap. Sensor dots: $\lambda=20$~\textmu eV; point contact: $\lambda=50$~\textmu eV in (a) and (c), $20$~\textmu eV in (b). (a) Local exponent of the optimised information rate. (b) Sensitivity gain of the sensor dot over the point contact at equal coupling; faded segments, here and in (a), mark points at which a detector fails the rate criterion, and the dotted line marks the working gap. (c) The rate criterion $\hbar(\Gup+\Gdn)/\Delta$, with the threshold 0.1 dotted.}
\label{fig:floor}
\end{figure}

To locate the floor, \Fig{fig:floor} holds the matrix elements fixed and varies the gap alone. We define the floor as the smallest gap above which the local exponent of the sensor dot, computed by centred finite differences on the logarithmic grid, stays within 0.1 of 2, with the crossing interpolated linearly in $\ln\Delta$. This is an operating criterion for the model tested, not a bound, and its value can move by about 20\% with the tolerance, the optimisation and the gap sampling. The floor lies at 17, 29 and 44~\textmu eV for $\Gamma=10$, 20 and 40~\textmu eV at 10~mK, at 52--77~\textmu eV at 50~mK and at 96--121~\textmu eV at 100~mK, set by scales of order $\Gamma$ and $10k_BT$, where $T$ is the electron temperature of the detector leads; the estimate $\max(1.5\Gamma,12k_BT)$ reproduces these floors only to within about 40\%.

As the figure shows, above the floor the rate criterion is satisfied everywhere at $\lambda=20$~\textmu eV (at $\lambda=50$~\textmu eV it fails below about 40--50~\textmu eV). Below the floor the local exponent falls smoothly, to about 1 at 10~mK and further at 100~mK, where the probe is thermally excited, and below about 10~\textmu eV the telegraph model is no longer controlled. The local exponent of the point contact is close to 1.1 around the working gap, where its back-action dominates. At smaller gaps it rises towards 1.8 at 10~mK, as the intrinsic relaxation takes over, whereas at 100~mK it falls below 1 once $k_BT$ becomes comparable with the gap. At equal coupling (20~\textmu eV) the sensitivity gain of the sensor dot over the point contact at the working gap is 3.1--3.6 at 100~mK and 5.8--9.6 at 10~mK for $\Gamma=10$--40~\textmu eV, and it grows as the gap closes, because in this comparison the sensor dot depends more steeply on the gap (local exponent 2) than the point contact at $\lambda=20$~\textmu eV (about 1.55 near the working gap), until the floor is reached.

\textit{Temperature.} The electron temperature enters in three ways, and 100~mK, a common base electron temperature in GaAs devices, is where they begin to matter at the gaps studied. First, the thermal tails of the lead distributions let a sensor dot biased below the gap emit at the gap with a weight of order $\mathrm{e}^{-(\Delta-eV)/k_BT}$; at the working point this gives an excitation rate of $1.3\times10^{2}$~s$^{-1}$ at 100~mK (and none that the calculation can resolve at 10~mK), still negligible, but at a gap of 100~\textmu eV and the same bias the exponent is only $-4$ and the thermal excitation competes with the measurement. This is what sets the 100~mK floor at 96--121~\textmu eV, about $12k_BT$, against $\Gamma$ at 10~mK. Second, thermal broadening of the lead distributions lowers the gain of a sensor dot whose width is comparable with $k_BT$ (8.6~\textmu eV at 100~mK against $\Gamma=20$~\textmu eV), which is why its best sensitivity at the working gap is 2.2 times worse at 100~mK than at 10~mK (\Tab{tab:matched}) and its optimum bias moves from 24 to 66~\textmu V. Third, the intrinsic bath populates $|f\rangle$ thermally through $n_B$, which is $10^{-9}$ at the working gap and 100~mK and matters only below the floor. For the point contact the same tails mean that the emission threshold is no longer sharp, and its local exponent falls below 1 once $k_BT$ becomes comparable with the gap (\Fig{fig:floor}(a)). The practical consequence for the Coulomb-type probe of \Sec{sec:robust}, whose working gap of 73~\textmu eV lies below the 100~mK floor, is that its inverse-square regime needs an electron temperature of about 50~mK or below.

\section{The interacting pair through a detector}\label{sec:pair}

\TAB{tab:matched} compares the pair with the matched single electron read through the same detector, with the same rate model, calibration and $T_1$. The ranking follows from \Eq{eq:transduction}: the transduced signal scales as $\Xf^{2}/\Delta$ and the information as $\Xf^{4}/\Delta^{2}$, whereas the QFI scales as $\Xf^{2}/\Delta^{2}$. The dipole advantage of the pair, $7.31/5.65=1.29$, therefore enters to the fourth power, 2.8, and with the detector kernel the static ratio is 2.99. Through the point contact, back-action lowers this to 2.5--2.9 in the two-level model; both probes are excited at similar rates, so most of the loss cancels in the ratio. The higher states lower it further, to 2.1--2.7, because the single electron's next level (0.45~meV, against 0.55~meV for the pair) is reached at a lower bias, and this is the value we quote (\App{app:multilevel}). Through the sensor dot the pair keeps the full static ratio of 2.99 at every coupling and temperature tested, because the excitation of either probe is negligible. The bias and level are optimised separately for each probe, so the comparison is at equal detector parameters, not at equal bias.

What the advantage is, and is not, should be stated precisely. It is a ratio of information rates for two probes with the same gap, read by the same detector at the same coupling and temperature, in the weak-coupling regime where the detector is limited by its own imprecision; it comes from the transition dipole, which enters the transduced signal twice. It holds for the sensor dot wherever the probe stays in its ground state, and for the point contact only to the extent that both probes are excited similarly. It is not a resource-independent quantum advantage: an ideal projective measurement of the probe state would restore the ranking of the QFI, 1.67, and a comparison at equal gate voltage rather than equal gap, or with a detector that couples to the two probes with different geometry, would give a different number. The two benchmarks also rank two independent single-electron devices differently: in the QFI the pair holds $1.67/2=0.84$ of their information, whereas through this readout it holds $2.99/2=1.5$ times their information rate.

\begin{table}[t]
\caption{Matched comparison at the working gap ($\Delta=181$~\textmu eV): ratio of the mean-current information rate of the pair to that of the single electron with the same gap and the same detector, and the sensitivity of the pair, $\Fm^{-1/2}$ in V\,m$^{-1}$\,Hz$^{-1/2}$. The coupling $\lambda=5$~\textmu eV is the level shift per interdot transfer measured for a sensor dot in~\cite{Haldar2025}. QFI ratio 1.67; static transduction ratio 2.99; sensor dot with $\Gamma=20$~\textmu eV; $T_1=16$~ns for both probes. The point-contact values are for the effective spectrum of \Sec{sec:detectors} and the two-level description; \Tab{tab:multilevel} gives them with the higher states of the probe included, and \Tab{tab:spectra} for two other spectra. Sensitivities use the two-sided convention of \App{app:detectors}; the one-sided noise-equivalent field is larger by $\sqrt2$.}
\label{tab:matched}
\centering
\begin{tabular}{llcccccc}
\toprule
 & & \multicolumn{2}{c}{$\lambda=5$~\textmu eV} & \multicolumn{2}{c}{$\lambda=20$~\textmu eV} & \multicolumn{2}{c}{$\lambda=50$~\textmu eV}\\
\cmidrule(lr){3-4}\cmidrule(lr){5-6}\cmidrule(lr){7-8}
Detector & $T$ & Ratio & Sensitivity & Ratio & Sensitivity & Ratio & Sensitivity\\
\midrule
Point contact & 10~mK & 2.79 & 0.50 & 2.48 & 0.18 & 2.90 & 0.084\\
Point contact & 100~mK & 2.79 & 0.52 & 2.47 & 0.19 & 2.87 & 0.088\\
Sensor dot & 10~mK & 2.99 & 0.095 & 2.99 & 0.024 & 2.99 & 0.0095\\
Sensor dot & 100~mK & 2.99 & 0.21 & 2.99 & 0.053 & 2.99 & 0.021\\
\bottomrule
\end{tabular}
\end{table}

\begin{figure}[t]
\centering
\includegraphics{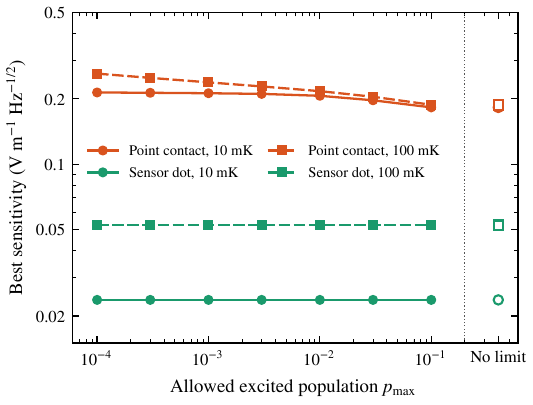}
\caption{Best field sensitivity of each detector when the excited population of the probe is constrained to $p_f\le p_{\max}$, at the working gap, with $\lambda=\Gamma=20$~\textmu eV, $T_1=16$~ns and a bias ceiling of 1~mV. Open symbols: unconstrained optima. The bias of the point contact is optimised within the allowed interval; the constraint on the sensor dot is inactive over the range shown.}
\label{fig:constrained}
\end{figure}

Equal coupling is one fair comparison; an equal limit on excitation is another. \FIG{fig:constrained} shows the best sensitivity of each detector when the excited population of the probe must stay below $p_{\max}$, with the bias (and the level of the sensor dot) re-optimised under that constraint. As the figure shows, the constraint binds only the point contact, whose optimum bias falls from 0.46~mV to 0.20~mV at $p_{\max}=10^{-2}$ and to 0.18~mV (10~mK) or 0.16~mV (100~mK) at $10^{-3}$; the population left by the sensor dot (about $10^{-6}$ or below at 100~mK) lies below every limit shown. The sensitivity gain of the sensor dot therefore rises as the limit tightens, from 7.7 without the constraint to 8.7 at $10^{-2}$ and 9.0 at $10^{-3}$ at 10~mK, and from 3.6 to 4.1 and 4.5 at 100~mK. The unconstrained points of \Fig{fig:constrained} are the $\lambda=20$~\textmu eV entries of \Tab{tab:matched}: the gain of 7.7 at 10~mK is the ratio of the pair sensitivities 0.18 and 0.024 in the table (7.7 before rounding), and 3.6 at 100~mK is 0.19 against 0.053. The table compares the two probes through one detector, the figure the two detectors for one probe, and both report the same optimised information rates. An excitation limit does not bound dephasing, static renormalisation or heating, so this is a comparison under one kind of disturbance.

The working point is also narrow in the confinement range. For the sensor dot at fixed bias and level ($\lambda=\Gamma=20$~\textmu eV, 100~mK) the sensitivity stays within 20\% of its optimum only for $8.0275\le R\le8.040$ (on a $0.0025\,a^{*}$ grid), a window of $\pm0.08\%$ in $R$. This number has no device-independent meaning, since $R$ is set by gate voltages through a lever arm that varies from device to device; what an experiment controls and measures is the gap. Across the window the odd gap moves from 212 to 188~\textmu eV through its minimum of 181~\textmu eV and the dipole from 7.24 to 6.89~$a^{*}$, so keeping the sensitivity within 20\% of its optimum means keeping the gap within about 15\%, or 30~\textmu eV, of its minimum. Level splittings in gate-defined double dots are set and held at the microelectronvolt level by gate detuning~\cite{Petta2004,Haldar2025}, the gate-to-energy conversion being the lever arm that each device must calibrate, and the working point can be found in situ by maximising the transduced signal against the gate, so the window is narrow but not beyond practice; a slow drift of the gates would show up as a drift of the gap and can be corrected by the same sweep.

\section{Discussion}\label{sec:discussion}

\subsection{Experimental predictions}\label{sec:predictions}

For a charge-sensed few-electron dot whose avoided-crossing gap can be tuned by a gate, the results make four predictions that a bias and gate sweep can test once the coupling and noise of the device are calibrated. (1) With a point contact whose back-action dominates the relaxation, the sensitivity (the smallest detectable field per root hertz) scales as $\Delta^{1/2}$, not as $\Delta$: halving the gap improves it by $\sqrt2$. With a sensor dot biased below the gap and its level held near a lead chemical potential, it scales as $\Delta$ down to the floor, of order $\Gamma$ and $10k_BT$. On logarithmic axes the two detectors give slopes of $1/2$ and 1, provided $\mu\Xf$, $\lambda$, the detector settings and the intrinsic relaxation stay fixed along the sweep; otherwise the slope of the calibrated quantity $|\mu\Xf|\Fm^{-1/2}$ should be compared. In our interaction scan, where $\mu\Xf$ grows as the gap closes, the uncorrected exponents of $\Fm$ are 1.38 and 1.35 for the point contact and 2.20 for the sensor dot at 10~mK. (2) The optimum bias of the point contact moves with the gap, at or above the emission threshold, whereas that of the sensor dot stays below the gap, at a value set by its level width. (3) A turnover of the sensitivity with bias tests the competition between detector-induced and intrinsic relaxation: for a point contact with flat transmission at half transmission and zero temperature, neglecting the telegraph and population-derivative terms, it lies at $eV^{*}=\Delta+2\hbar\Gamma_0/(\lambda^2\mu^2K)$, with $K=\pi/[2(\hbar\omega_x)^2]$ the back-action normalisation of \App{app:detectors}, which gives 280~\textmu eV at the working point for $\lambda=50$~\textmu eV. The full two-level model places the optimum at 250~\textmu eV, and the difference comes from the two terms the formula drops: the telegraph noise grows with the excited population and penalises bias beyond the threshold, and the population-derivative term $\delta\,\partial_\varepsilon p_f$ adds signal that also peaks near the threshold, so both pull the optimum down; the energy dependence of the transmission, which lowers the gain at higher bias, moves it in the same direction. With the higher states of the probe included the optimum moves back up, to 320--340~\textmu eV (\App{app:multilevel}): at these biases the third level is still not populated, but the field mixes $|f\rangle$ with it, which adds a term $2\varepsilon X_{f2}M_{f2}/(E_f-E_2)$ to the signal carried by the excited state and so reduces the cost of populating it. The formula is therefore a guide to the scale of the turnover, within about 20\%, not a prediction of its position. A weak detector pushes the turnover beyond the usable bias window, and a strong one pins it near the threshold. (4) A readout that is even under parity, such as the quantum capacitance seen by a gate placed symmetrically about the dot centre, shows no first-order response at the symmetric point; a detector on one side is required.

The platform exists. Haldar \textit{et al}~\cite{Haldar2025} monitored a double-dot two-level system with a sensor dot and measured the decoherence it imposes, proportional to the tunnelling rate of the detector and about twice their model prediction. Their detector operated at millivolt biases, well above the splitting of a few tens of microelectronvolts of their two-level system, and their measured level shift per interdot transfer, about 5~\textmu eV, calibrates the coupling scale of \Tab{tab:matched}. That experiment did not explore the regime below threshold. Its architecture offers a route to testing it, provided the gap, the detector width and the electron temperature are tuned to give substantial measurement gain below the probe gap; lowering the bias alone does not create this separation of scales, since in that device the detector width already exceeds the splitting. Horibe \textit{et al}~\cite{Horibe2015} found operating conditions of a single-electron-transistor detector with both good sensitivity and back-action-induced excitation rates below a kilohertz. Our results assume second-order (golden-rule) rates, a detector memory short compared with the probe dynamics, secular separation of the rates from the gap, and linear detector response, $|\lambda\delta|\lesssim\max(\Gamma,k_BT)$; the last condition concerns only the diagonal contrast and does not by itself control the transverse coupling $\lambda\mu$. The independence of the exponent from $\lambda$ is established within these assumptions, not for arbitrary coupling.

\subsection{Robustness to the form of the interaction}\label{sec:robust}

The contact interaction stands in for the screened repulsion of a quasi-one-dimensional dot. We have repeated the calculation with the Coulomb interaction averaged over a Gaussian transverse profile of width $\ell=0.5$--$2\,a^{*}$, and with the soft-Coulomb form $(x^2+b^2)^{-1/2}$ at $b=1\,a^{*}$. The crossover survives and sharpens. Taking as working point the largest transition dipole subject to a gap of at least $5\times10^{-3}\,\Ha$, it lies at $R=8.039$, with a gap of 71--73~\textmu eV and $\Xf=9.22$--9.25~$a^{*}$, larger than $d$ because the repulsion pushes the electrons outwards; there $\mu/\Xf=0.065$, close to $1/2d=0.0625$, and $\delta/2\mu=-0.01$ to $-0.02$ (\App{app:ci}). The matched single electron at this gap has $X_{01}=7.20\,a^{*}$, so the QFI ratio is 1.64 and the static transduction ratio 2.96--2.97, against 1.67 and 2.99 for the contact interaction. The advantage of the pair is therefore similar for the interactions tested; what changes is the gap of the working point.

At 73~\textmu eV the Coulomb working point lies above the floors of \Sec{sec:floor} at 10~mK for all three widths and at 50~mK for $\Gamma\le20$~\textmu eV, but below them at 100~mK; these floors were computed with the matrix elements of the contact probe and are only an estimate for the Coulomb probe ($\mu=0.60$, $\delta=-0.02$). Through a sensor dot with $\Gamma=\lambda=20$~\textmu eV the pair-to-single ratio is 2.97 at 10 and 50~mK and 2.96 at 100~mK, where the excited population of the pair reaches 1.5\%: the ratio survives, but the inverse-square regime is not established at 100~mK for this probe.

\subsection{Dependence on the detector model and its parameters}\label{sec:parameters}

The reported improvements rest on a detector model with several chosen quantities, and we have varied each of them. Two conclusions are robust and one number is not.

\textit{The spectrum of the point contact.} \TAB{tab:spectra} repeats the point-contact calculation with three back-action spectra: the Aguado--Kouwenhoven form used throughout (A), an effective variant in which the non-symmetrised current noise of the same saddle-point contact is evaluated with its energy-dependent transmission, the reflection phases being taken equal (B), and the partition noise alone, without the equilibrium term (C); all three are normalised by the same quantum-limit condition and are defined in \App{app:detectors}. Across the three forms, the couplings 20 and 50~\textmu eV, the temperatures 10 and 100~mK and the widths 10--40~\textmu eV, the sensitivity gain of the sensor dot over the point contact at the working gap lies between 2.4 and 11, against 3.1--11 for form A alone over the same set (\Tab{tab:matched} extends the lower end to 2.5 at $\lambda=5$~\textmu eV, for which only form A has been run), and its optimum bias stays below the gap whatever the point contact does. The fitted exponent of the point contact lies between 0.95 and 1.17 for the three forms (fits over the eight basis-extrapolated gaps of \Sec{sec:test}), always far from the 2.00 of the sensor dot. The only number that moves by more than the quoted ranges is the zero-temperature, detector-dominated exponent of form B, 0.70 instead of 1.02: with the energy-dependent transmission the spectrum is no longer homogeneous in the bias, since $\hbar\omega_x$ is a second scale, and the optimum bias runs into the 1~mV ceiling at the larger gaps. This is the situation anticipated in \Sec{sec:single}: the inverse-gap law is the homogeneous-spectrum limit, and a real contact whose transmission varies on the scale of the bias gives an exponent between about 0.7 and 1.2.

\begin{table}[t]
\caption{Point contact with three back-action spectra at the working gap (two-level model, bias ceiling 1~mV). Sens.: best sensitivity of the pair in V\,m$^{-1}$\,Hz$^{-1/2}$; $eV^{*}$: optimum bias; gain: sensitivity gain of a sensor dot with $\Gamma=20$~\textmu eV at the same coupling and temperature; exponent: fitted gap exponent of the point contact over the eight basis-extrapolated gaps of \Sec{sec:test} at $\lambda=50$~\textmu eV, with the zero-temperature detector-dominated value in brackets.}
\label{tab:spectra}
\centering
\small
\setlength{\tabcolsep}{4pt}
\begin{tabular}{llccccccc}
\toprule
 & & \multicolumn{3}{c}{$\lambda=20$~\textmu eV} & \multicolumn{3}{c}{$\lambda=50$~\textmu eV} & \\
\cmidrule(lr){3-5}\cmidrule(lr){6-8}
Spectrum & $T$ & Sens. & $eV^{*}$ (\textmu V) & Gain & Sens. & $eV^{*}$ (\textmu V) & Gain & Exponent\\
\midrule
A & 10~mK & 0.183 & 456 & 7.7 & 0.084 & 232 & 8.9 & 1.17 (1.02)\\
A & 100~mK & 0.188 & 465 & 3.6 & 0.088 & 250 & 4.2 & 1.14\\
B & 10~mK & 0.143 & 1000 & 6.0 & 0.064 & 1000 & 6.8 & 0.95 (0.70)\\
B & 100~mK & 0.146 & 1000 & 2.8 & 0.066 & 1000 & 3.1 & 0.95\\
C & 10~mK & 0.191 & 398 & 8.0 & 0.085 & 182 & 9.0 & 1.06 (1.02)\\
C & 100~mK & 0.196 & 409 & 3.7 & 0.091 & 195 & 4.3 & 1.04\\
\bottomrule
\end{tabular}
\end{table}

\textit{Coupling, width, intrinsic relaxation and bias ceiling.} The coupling $\lambda$ changes the absolute sensitivities as $1/\lambda$ for the sensor dot and the pair-to-single ratio of the point contact by 15\% (\Tab{tab:matched}), but not the exponents, which are the same at 20 and 50~\textmu eV for the sensor dot. The width $\Gamma$ sets the floor and the absolute sensitivity of the sensor dot ($\Fm\propto1/\Gamma$ above the floor) and changes the gain over the point contact by a factor of 1.7 between 10 and 40~\textmu eV (\Fig{fig:floor}(b)). The intrinsic relaxation time $T_1$ decides whether the point contact is dominated by its own back-action: with $T_1=16$~ns this is the case at $\lambda=50$~\textmu eV but not at 20~\textmu eV, where the exponent moves to about 1.6 (\Sec{sec:test}); a shorter $T_1$ would move the crossover to larger couplings and leave the sensor dot unaffected until $\Gamma_0$ competes with its measurement rate. The bias ceiling of 1~mV binds only the point contact: for form B at $\lambda=20$ and 50~\textmu eV, and for form A at $\lambda=5$~\textmu eV.

\textit{The probe.} The interaction form changes the gap of the working point but not the conclusions (\Sec{sec:robust}). The confinement parameters $V_0$, $d$ and $p$ set the position and sharpness of the crossover; for $p\gtrsim100$ the sensing figures no longer depend on $p$ (\Sec{sec:probe}), and a different $d$ rescales the dipole with the well separation. The results that do not depend on any of these choices are the two statements of \Sec{sec:scaling}: a detector whose back-action spectrum has the bias as its only scale, and which dominates the relaxation, gives an exponent of 1 for the mean-current information, and a detector that reaches full measurement strength below the gap gives 2 above a floor set by its own energy scales. The size of the gain between the two detectors, 2.4--11 at couplings of 20--50~\textmu eV, is the number that depends on the model.

\subsection{Limits of the model}\label{sec:limits}

\textit{The probe.} The probe is one-dimensional; a two-dimensional gate geometry can change the gaps, the dipoles, the ordering of the states and the relaxation, and a two-dimensional configuration-interaction treatment remains to be done. Our protocol is the stationary state of the two-level probe under continuous readout, which is close to the ground state for the sensor dot; a confinement ramp that preserves parity drives transitions only in the even sector, whose gap does not close at the crossover, but the preparation time of a specified ramp is not evaluated here (see~\cite{Rams2018,Mihailescu2026gap} for the timing costs of gap-dependent protocols).

\textit{Higher states.} At the working point the next singlets lie 0.55, 0.76, 0.98, 1.19, 1.27 and 1.48~meV above the ground state, with $|M_{f2}|=0.37$ in transfer units. The sensor-dot operating points cannot reach them at zero temperature, but the point-contact optima can: at $\lambda=20$~\textmu eV the optimum bias of 0.46~mV allows sequential excitation from $|f\rangle$, and at $\lambda=5$~\textmu eV the optimum near 1~mV allows direct excitation of the third and fourth and sequential excitation of higher levels. \APP{app:multilevel} solves the rate equations with the lowest three, four, six and eight singlets. The higher levels are populated to at most $4\times10^{-3}$ at the point-contact optima. Between two and four levels the point-contact gap exponents change by at most 0.06; between two and eight levels its sensitivities change by about 6\% at most and its pair-to-single ratio falls from 2.5--2.9 to 2.1--2.7 (by up to 21\%); from four to eight levels the sensitivities and pair-to-single ratios change by less than 1\% and the optimum bias by at most 1.4\% (the exponents were not recomputed beyond four levels); the sensor dot is unchanged. The two-level numbers are therefore accurate to a few per cent for the point contact, within the rate model, and exact at this order for the sensor dot.

\textit{The detectors.} The sensor dot is a spinless, non-interacting level, without Kondo correlations, interaction-induced inelastic cotunnelling or the many-body correlations in the detector leads that can exchange energy with the probe~\cite{Grant2026}. Coulomb blockade, which the model also omits, is examined in \App{app:blockade} in the sequential-tunnelling regime $\Gamma\ll k_BT$ of a blockaded dot read through one charge transition, with and without spin degeneracy, at the working gap. In that regime energy filtering survives, because it rests on the lead distributions and not on the coherence of the level; the gain below the gap saturates once $eV$ exceeds a few $k_BT$ instead of $\Gamma$, the optimum bias stays below the gap, and the absolute sensitivity differs from that of a non-interacting level of the same width by 18--25\%, the spin-degenerate blockaded dot being the more sensitive. This is a demonstration for one regime, not a validation of the non-interacting model for a spin-degenerate interacting detector at $\Gamma\gtrsim k_BT$, which remains an assumption of the paper; what the examples show is that the two properties the laws rest on, measurement at full strength below the gap and the absence of emission above the bias, do not depend on that assumption. The back-action spectrum of the point contact has an assumed shape with one normalisation, because its intrinsic charge noise and its circuit-mediated current noise couple differently~\cite{Young2010}; \Sec{sec:parameters} gives the range of the results over three forms of the spectrum, each of which uses a current-noise spectrum as the back-action spectrum. The coupling $\lambda$ is the same for both detectors by assumption, so the comparison is one of energy structure rather than of geometry.

\textit{Phonons and radio-frequency readout.} In GaAs lateral dots the dominant back-action of a biased point contact is carried by non-equilibrium phonons~\cite{Gasser2009,Harbusch2010,Granger2012}. If that channel is homogeneous in the bias and dominates the transitions while the imprecision remains electrical, the inverse-gap law of the mean current survives with a different prefactor and optimum (\App{app:detectors}); we do not establish either condition for a real device. The suppression below threshold extends to phonons only if the emitted phonon spectrum obeys the same $eV$ cut-off, which holds for single hot electrons but is not established for correlated processes. For a radio-frequency sensor dot the carrier adds to the dc bias: keeping the dc bias plus the rf amplitude below $\Delta$ is only a quasistatic criterion, and photon-assisted processes at the carrier frequency need a calculation for a driven detector, which we have not made. The electron temperature of the detector leads sets the thermal floor and must be measured.

\textit{The telegraph description.} The telegraph description rests on three approximations, and \App{app:checks} tests each at the operating points of \Tab{tab:matched}; the tests are comparisons between second-order treatments at those points, not bounds on higher orders, on detector memory or on correlations between the output noise and the back-action, which remain untested. The secular approximation, which drops the coherences between $|0'\rangle$ and $|f'\rangle$, is tested against the non-secular Bloch--Redfield equation of the probe coupled to each detector's spectrum, with the Lamb shift in its secular form, at five of the operating points of \Tab{tab:matched} (sensor dot at $\lambda=20$~\textmu eV and 10 and 100~mK and at $\lambda=50$~\textmu eV and 10~mK; point contact at $\lambda=50$~\textmu eV and 100~mK and at $\lambda=20$~\textmu eV and 10~mK): the stationary populations (absolute) and the transduced signal (relative) agree to $10^{-7}$ with two and with three levels; the noise is not recomputed in that model. The static renormalisation of the probe by the detector, the second-order shift of its levels, is up to 6~\textmu eV for the sensor dot and below 1~\textmu eV for the point contact at the couplings studied, at most 3.5\% of the gap. The restriction to the mean current, which drops the information in the time structure of the record, is tested at four point-contact operating points, where no excess information is resolved within the 1--2\% numerical uncertainty, and bounded for the sensor dot, by about $10^{-6}$ at the working point and by $3\times10^{-4}$ along the 100~mK scan above the floor. Master equations with energy-dependent back-action, and their conditional versions, exist for a point contact at arbitrary bias~\cite{Stace2004,Stace2003}; a conditional model of this kind, consistent with the detector spectra used here and with the correlations between output noise and back-action, has not been implemented, and is what a direct test of the record would need. The Markov assumption, that the detector has no memory on the time scale of the probe's rates, is not tested; its parameter, the ratio of the rates to the detector's bandwidth ($\Gamma$ for the sensor dot, $eV$ and $\hbar\omega_x$ for the point contact), is below 0.02 at the operating points of \Tab{tab:matched} and below 0.04 along the sensor-dot scan at $\lambda=20$~\textmu eV, and reaches 0.24 only at the smallest gap of the scan at $\lambda=50$~\textmu eV; the rate criterion of \Sec{sec:test} is applied to all of them.

\textit{Other materials.} In effective atomic units the probe transfers to other hosts by rescaling; in silicon the working gap is about three times larger and the dot three times smaller than in GaAs, which makes the thermal floor easier to clear, whereas the detector parameters do not rescale and the valley states of silicon lie outside the model.

\section{Conclusions}\label{sec:conclusions}

A sensor near a closing gap is only as good as what its detector can see, and in a semiconductor device the detector that reads the gap can also excite the probe across it. We have shown, for two interacting electrons at a confinement-driven crossover in a gate-defined quantum dot, that the outcome is decided by the energy structure of the detector. A point contact must buy measurement strength with bias, and above the emission threshold the same bias excites the probe: when its back-action dominates, the closed-system enhancement $\Delta^{-2}$ becomes $\Delta^{-1}$ in the mean-current information rate, and, for the observation model of \App{app:detectors}, in that of the whole record. A sensor dot reaches its full measurement strength at a bias set by its own level width, so it can be operated below the gap, where, for the detector spectra considered and to second order in the coupling, detector-induced excitation vanishes at zero temperature and stays small when $\Delta-eV$ is several $k_BT$. The $\Delta^{-2}$ enhancement then returns, for any weak coupling, above a floor set by scales of order $\Gamma$ and $10k_BT$.

Through such a detector the interacting pair beats a matched single electron by the full transduction factor, about 3 in the mean-current information rate for contact and Coulomb-type interactions alike, against 1.7 and 1.6 in the QFI. At equal coupling of 20--50~\textmu eV the sensor dot also reads the field 2.4 to 11 times more precisely than a point contact, for the widths, temperatures and point-contact spectra studied. These statements hold within a weak-coupling rate description whose approximations we have tested at the operating points studied: the higher states of the probe change the point-contact sensitivities by up to 6\%, its exponents by up to 0.06 and its pair-to-single ratio by up to 21\%, and leave the sensor dot unchanged; the coherences dropped by the rate equations change the populations and the signal by less than $10^{-7}$, and no excess information in the full detector record is resolved within the 1--2\% precision of that check; a Coulomb-blockaded sensor dot in the sequential-tunnelling regime keeps its optimum below the gap, with an absolute sensitivity that differs from the non-interacting level's by 18--25\% at equal width; and extrapolating the basis of the probe moves the fitted exponents by at most 0.05. The two laws predict the slope of the sensitivity against the gap and the behaviour of the optimum bias, and both can be tested on suitably tuned charge-sensed double dots. For the design of gap-enhanced charge sensors the lesson is simple: a detector should be chosen for where its back-action sits in energy, not only for how large it is.

\section*{Acknowledgments}

The author thanks the Kurdistan Institution for Strategic Studies and Scientific Research for financial support, and the two anonymous referees for comments that led to the checks of \App{app:checks}. Generative AI tools (Claude, Anthropic, model Claude Fable 5.1) were used under the author's direction to assist with the numerical codes, the figures and the consistency of numbers between the text and the result files; all results were verified by the author, who takes full responsibility for the manuscript.

\section*{Data availability statement}

The data that support the findings of this study are openly available in the Zenodo record \href{https://doi.org/10.5281/zenodo.23108551}{10.5281/zenodo.23108551}; the scripts and data added in revision (appendices~A and C) are deposited as a new version of the same record. The archive contains the scripts, the data for the figures and tables and the validation outputs described in \App{app:ci}, with a README, a dependency list and run logs.

\appendix
\titleformat{\section}{\normalfont\large\bfseries}{Appendix \thesection.}{0.6em}{}

\section{Configuration-interaction solver and convergence}\label{app:ci}

Single-particle orbitals are obtained on a real-space grid ($|x|\le26$, spacing 0.01) by diagonalising the tridiagonal kinetic-plus-potential matrix. The lowest $n_{\mathrm{orb}}$ orbitals span the two-electron singlet space of dimension $n_{\mathrm{orb}}(n_{\mathrm{orb}}+1)/2$, in which the contact interaction is evaluated by quadrature of the pair densities. For a long-range interaction the singlet matrix element is $\langle ab|U|cd\rangle=n_{ab}n_{cd}[J_{ac,bd}+J_{ad,bc}]$, with $J_{ij,kl}=\iint\phi_i\phi_j(x)\,U(x-y)\,\phi_k\phi_l(y)\,\mathrm{d}x\,\mathrm{d}y$ and $n_{ab}=(1+\delta_{ab})^{-1/2}$ (direct and exchange terms); the $J$ integrals are evaluated by convolving each pair density with the interaction on the grid. Detector operators are built from the orbital matrix of $f(x)$ and transformed to the eigenbasis. The field-coupled state used for the working points and scans is the excited singlet with the largest $|X_{0n}|$ among the lowest eight; \Fig{fig:crossover} instead follows the lowest odd state by parity, which coincides with it between $R=8.00$ and 8.07.

\textit{Contact interaction: convergence.} The cusp of the contact interaction makes the orbital expansion converge algebraically, and the more slowly the stronger the interaction. \TAB{tab:convergence} gives, for each working point of the interaction scan, the gap with 50, 70, 90, 110, 130 and 150 orbitals at fixed $R^{*}(g)$, together with two extrapolations to the complete basis: linear in $1/n_{\mathrm{orb}}$ over the three largest bases, which is the form the last increments follow (the increments per 20 orbitals at $g=8$ are 2.3, 2.9, 2.8, 2.5 and 2.2~\textmu eV, consistent with a $1/n_{\mathrm{orb}}$ tail), and quadratic in $1/n_{\mathrm{orb}}$ over the four largest. The two forms agree to within 3\% for $g\le3$ and differ by 5\%, 11\% and 16\% at $g=4$, 6 and 8, and this difference, not the much smaller spread between fit windows of one form (2~\textmu eV at $g=8$), is the uncertainty we attach to the extrapolated gaps: $62.7\pm7$~\textmu eV at $g=6$ and $49.5\pm8$~\textmu eV at $g=8$. The available bases do not certify the asymptotic form, so these are basis-extrapolated estimates; a 200-orbital calculation, which would provide a further check, exceeded the memory available to us. The 150-orbital gaps themselves lie 21\% and 26\% below the linear extrapolation at $g=6$ and 8, so the extrapolation is a large correction there. At the working point ($g=1$) the 50-orbital gap is low by 2.3\% and the dipole high by 0.9\%, with the two forms agreeing to 0.3\%; these are the values used in sections~\ref{sec:probe}--\ref{sec:detectors} and \ref{sec:pair}, and their basis error is smaller than the model uncertainties quoted there. Along the scan the error of the 50-orbital gap grows to 44\% and 52\% at $g=6$ and 8, with the dipole converging faster (5\% and 4\%); the detector elements $\mu$ and $\delta$ move by up to 5\% and a factor of two, the latter because $\delta$ is a small difference of two large elements. The matrix elements converge from above and the gaps from below, so a 50-orbital calculation overestimates $4\Xf^2/\Delta^2$ at strong interaction. Relocating the working point with 90 orbitals moves $R^{*}$ by 0.012 at $g=6$ and changes the gap and the dipole there by 1.2\% and 0.5\%, far less than the basis effect at fixed $R$, so the working points are kept at their 50-orbital positions and their properties are extrapolated. \SEC{sec:test} and \Fig{fig:scaling} use the linear extrapolation and report the exponents for the 50-orbital, 150-orbital, linear and quadratic sets; the sensor-dot exponents do not depend on the basis except through the membership of points near the 100~mK floor, and the point-contact exponent at 10~mK moves from 1.22 to 1.17. The solver also reproduces qualitatively the entanglement crossover of~\cite{Abdullah2009} (a double-well plateau of the ground-state linear entropy near 0.5 and a small minimum on the single-well side).

\begin{table}[t]
\caption{Basis convergence of the working points of the interaction scan (contact interaction, $R^{*}(g)$ located with 50 orbitals and held fixed). Gap in \textmu eV with 50 to 150 orbitals and two extrapolations to $n_{\mathrm{orb}}\to\infty$: linear in $1/n_{\mathrm{orb}}$ over the three largest bases (lin., the values used in \Sec{sec:test}) and quadratic in $1/n_{\mathrm{orb}}$ over the four largest (quad.); the difference between the two is the uncertainty of the extrapolated gap quoted in the text. $\Xf$ (in $a^{*}$) with 50 orbitals and both extrapolations; $\mu$ and $\delta$ (in transfer units) with 50 orbitals and the linear extrapolation.}
\label{tab:convergence}
\centering
\footnotesize
\setlength{\tabcolsep}{3pt}
\begin{tabular}{cccccccccccccc}
\toprule
$g$ & $\Delta_{50}$ & $\Delta_{70}$ & $\Delta_{90}$ & $\Delta_{110}$ & $\Delta_{130}$ & $\Delta_{150}$ & $\Delta_\infty^{\mathrm{lin.}}$ & $\Delta_\infty^{\mathrm{quad.}}$ & $\Xf^{(50)}$ & $\Xf^{\mathrm{lin.}}$ & $\Xf^{\mathrm{quad.}}$ & $\mu^{(50)}\to\mu^{\mathrm{lin.}}$ & $\delta^{(50)}\to\delta^{\mathrm{lin.}}$\\
\midrule
0.7 & 205.6 & 205.9 & 206.2 & 206.4 & 206.6 & 206.7 & 207.5 & 207.8 & 7.228 & 7.201 & 7.197 & $0.457\to0.455$ & $0.105\to0.107$\\
1 & 181.6 & 182.3 & 182.9 & 183.4 & 183.8 & 184.1 & 186.0 & 186.5 & 7.328 & 7.262 & 7.255 & $0.462\to0.458$ & $0.109\to0.113$\\
1.5 & 145.9 & 147.1 & 148.3 & 149.2 & 149.9 & 150.5 & 154.1 & 155.1 & 7.499 & 7.374 & 7.359 & $0.472\to0.463$ & $0.103\to0.110$\\
2 & 118.6 & 120.3 & 121.9 & 123.3 & 124.3 & 125.1 & 130.2 & 131.7 & 7.633 & 7.455 & 7.433 & $0.480\to0.468$ & $0.092\to0.102$\\
3 & 82.1 & 84.4 & 86.8 & 88.8 & 90.4 & 91.6 & 99.4 & 102.0 & 7.810 & 7.551 & 7.512 & $0.490\to0.472$ & $0.070\to0.086$\\
4 & 59.5 & 62.3 & 65.3 & 67.8 & 69.8 & 71.5 & 81.7 & 85.7 & 7.924 & 7.604 & 7.546 & $0.497\to0.475$ & $0.053\to0.074$\\
6 & 35.1 & 37.9 & 41.2 & 44.3 & 47.0 & 49.2 & 62.7 & 69.9 & 8.071 & 7.695 & 7.596 & $0.506\to0.480$ & $0.032\to0.059$\\
8 & 23.9 & 26.2 & 29.1 & 31.9 & 34.4 & 36.6 & 49.5 & 57.7 & 8.162 & 7.824 & 7.718 & $0.512\to0.489$ & $0.022\to0.047$\\
\bottomrule
\end{tabular}
\end{table}

\textit{Long-range interactions.} The implementation is validated by tests with pass/fail tolerances, applied to the matrix that the solver actually assembles: a constant interaction returns the identity matrix to $10^{-15}$, a narrow Gaussian interaction reproduces the contact gap with a residual of $7\times10^{-5}$, and a four-orbital basis agrees with direct two-coordinate quadrature of explicitly symmetrised states with a residual of $1.3\times10^{-5}$ (acceptance thresholds $10^{-3}$). For the soft-Coulomb interaction with $b=1$ at $R=8.039$ the gap is 73.09, 73.08, 73.08 and 73.08~\textmu eV and $\Xf=9.2197$, 9.2196, 9.2196 and 9.2196 at 10, 20, 40 and 70 orbitals: without a cusp the orbital expansion converges by 20 orbitals, and halving the grid spacing at 20 orbitals changes the gap by $+0.016\%$. The quasi-one-dimensional interaction is $U(x)=\sqrt{\pi/2}\,\ell^{-1}\exp(x^2/2\ell^2)\,\mathrm{erfc}(|x|/\sqrt2\ell)$, which tends to $1/|x|$ at long range; its working points (largest dipole subject to $\Delta\ge5\times10^{-3}\,\Ha$, on a grid of $0.001\,a^{*}$) lie at $R=8.039$, with gaps of 73, 73 and 71~\textmu eV for $\ell=0.5$, 1 and 2. The transverse excitation energy of a dot of width $\ell=1$ is of order $1\,\Ha=12$~meV, far above every detector bias considered, so the detector cannot excite transverse states. The 41-point crossover scan of \Fig{fig:crossover} uses 50 orbitals, which is adequate at $g=1$ (\Tab{tab:convergence}); the interaction scan uses the extrapolated values of \Tab{tab:convergence}.

\section{Detector models}\label{app:detectors}

Energies are in meV and $\hbar=1$ inside the formulas; rates are converted with $1~\mathrm{meV}/\hbar=1.52\times10^{12}$~s$^{-1}$.

\textit{Point contact.} The point contact has transmission $\Tr(E)=[1+\mathrm{e}^{-2\pi(E-E_0)/\hbar\omega_x}]^{-1}$ with $\hbar\omega_x=1$~meV and $\tau=\Tr(0)=1/2$. Its current $I=\int\frac{\mathrm{d}E}{2\pi}\Tr(E)(f_L-f_R)$, gain magnitude $|\partial I/\partial\epsilon_0|$ (only its square enters $\Fm$) and zero-frequency noise $S_{II}=\int\frac{\mathrm{d}E}{2\pi}[\Tr(f_L(1-f_L)+f_R(1-f_R))+\Tr(1-\Tr)(f_L-f_R)^2]$ are integrated over the bias window (Landauer--B\"uttiker). For a flat transmission these reduce to $\partial I/\partial\epsilon_0=eV\Tr'/2\pi$, with $\Tr'=(2\pi/\hbar\omega_x)\tau(1-\tau)$, and $S_{II}=[2k_BT\tau^2+\tau(1-\tau)eV\coth(eV/2k_BT)]/2\pi$; the integrated gain falls below the flat value by 16\% at $eV=0.5$~meV and by 40\% at 1~meV. The back-action spectrum is $S(E)=K\{\tau(1-\tau)\sum_\pm g(E\pm eV)+2\tau^2g(E)\}$, with $g(x)=x/(1-\mathrm{e}^{-x/k_BT})$~\cite{Aguado2000} and $K=\pi/[2(\hbar\omega_x)^2]$, which makes $\Gmeas=\Gamma_\phi$ at small bias and zero temperature; the rates of \Eq{eq:rates} evaluate it at $E=\mp\Delta$. With the transmission integrated over the window the efficiency is $\eta=1$ only at small bias and zero temperature; at the optimised operating points of \Tab{tab:matched} it is 0.54--0.94 (0.58, 0.85 and 0.94 for the pair at 10~mK and $\lambda=5$, 20 and 50~\textmu eV).

\textit{Sensor dot.} The sensor dot is a resonant level with spectral function $A(E)=\Gamma/[(E-\epsilon_0)^2+\Gamma^2/4]$, transmission $\Tr=\Gamma_L\Gamma_RA/\Gamma$ with $\Gamma_L=\Gamma_R=\Gamma/2$, and lead-weighted distribution $n(E)=[f_L(E)+f_R(E)]/2$ at $\mu_{L,R}=\pm eV/2$. Its current, noise and gain follow from the same Landauer--B\"uttiker formulas with the energy-dependent $\Tr$, and its charge-noise spectrum is $S(E')=\int\frac{\mathrm{d}E}{2\pi}A(E)n(E)A(E+E')[1-n(E+E')]$, exact for a non-interacting level.

\textit{Alternative spectra for the point contact.} The two other spectra of \Tab{tab:spectra} are defined as follows; like form A, both use a current-noise spectrum of the contact as its back-action spectrum, which is an effective choice (\Sec{sec:limits}). Form B is the non-symmetrised current noise of the saddle-point contact with its energy-dependent transmission and reflection, $R=1-\Tr$,
\begin{align*}
S_B(E)=K_B\!\int\!\mathrm{d}E'\Big\{&\big[1-\sqrt{R(E')R(E'+E)}\big]^{2}f_L(1-f_L')+R(E')\Tr(E'+E)f_L(1-f_R')\\&+\Tr(E')R(E'+E)f_R(1-f_L')+\Tr(E')\Tr(E'+E)f_R(1-f_R')\Big\},
\end{align*}
with $f'=f(E'+E)$; the transmission probabilities do not fix the finite-frequency scattering phases, and the reflection phases are taken equal at $E'$ and $E'+E$, so this is an effective variant rather than the unique spectrum of the contact. For a flat transmission it reduces to form A. Form C is the partition noise alone, $S_C(E)=K\tau(1-\tau)\sum_\pm g(E\pm eV)$. $K_B$ and $K$ are fixed by $\eta=1$ at small bias and zero temperature. The emission side of all three vanishes for $|E|>eV$ at zero temperature.

\textit{Optimisation.} The operating points of \Figs{fig:bias}{fig:scaling} and of the tables are found as follows. For the point contact the scattering energy is fixed at $\tau=1/2$ and the bias is optimised by a bounded one-dimensional search of $\ln\Fm$ between $0.05\Delta$ and the ceiling of 1~mV. For the sensor dot the bias and the level position are optimised together: a grid of 22 biases, logarithmic between $0.3(\Gamma+k_BT)$ and $\min(20\Delta+10(\Gamma+k_BT),\,1~\mathrm{mV})$, and 25 level positions across the bias window plus $3(\Gamma+k_BT)$ on each side, followed by two rounds of local refinement (factors 0.8--1.25 in bias and steps of $0.2$--$0.5(\Gamma+k_BT)$ in level). Both optima are flat in the bias once the level of the sensor dot is re-optimised at each bias: at the working point and 100~mK, $\Fm$ stays within 10\% of its maximum between 45 and 160~\textmu V for the sensor dot (\Fig{fig:bias}) and between 180 and 400~\textmu V for the point contact, so the optimised information rates are insensitive to the grid, whereas the quoted optimum biases are uncertain by about a factor of 1.5 for the sensor dot and by $\pm30\%$ for the point contact. The sensor dot is more sensitive to its level position at fixed bias: $\pm5$~\textmu eV costs 6\% at 100~mK and 35\% at 10~mK, where the level must sit within $k_BT$ of the flank of its own resonance; this is the quantity an experiment tunes with the plunger gate. The uncertainties of the fitted exponents are the standard errors of the least-squares slopes (\Sec{sec:test}); those of the floors are the $\pm20\%$ of \Sec{sec:floor}.

\textit{Conventions.} The field sensitivity quoted throughout is $\Fm^{-1/2}$, with $S_q$ the two-sided covariance intensity of the imprecision; the corresponding one-sided noise-equivalent field is larger by $\sqrt2$. With output levels $\pm\delta/2$ and covariance intensity $S_q$, the overlap of the likelihoods of the integrated output decays as $\exp(-\delta^2t/8S_q)$, which defines $\Gmeas=\delta^2/8S_q$~\cite{Korotkov1999,Gurvitz1997}. The quantum limit $\eta=\Gmeas/\Gamma_\phi\le1$ is satisfied by both detectors, with $\eta\to1$ for the point contact at small bias and zero temperature, and $\eta=0.997$ for the resonant level at small bias. The optimisation over bias is restricted to $V\le1$~mV.

\textit{Scaling of the full record.} Take the two-state telegraph process of \Eq{eq:Fmean}, with the field entering through the level shifts $\pm a\varepsilon$ and through $M_{0'f'}$, and a record $I(t)$ formed by the state-dependent level plus Gaussian white imprecision noise of intensity $S_q$, uncorrelated with the back-action. Under assumptions (i)--(iv) of \Sec{sec:single}, at fixed $v=eV/\Delta$ and fixed dimensionless parameters ($\eta$, $\delta/\mu$, $\mu/\Xf$), every rate is proportional to $\Delta$, $S_q$ to $1/\Delta$, and every field sensitivity to $1/\Delta$. Rescaling time by $\Delta$ and the field by $1/\Delta$ then maps the likelihood of the record at one gap onto that at any other, so the Fisher information rate of the complete record, and of any statistic whose definition survives this rescaling, has the form $a^2\Delta\,\Phi(v,\eta,\delta/\mu,\dots)$ with $\Phi$ independent of $\Delta$. For this observation model the inverse-gap law is a property of the record, and the mean current attains it up to the $\Delta$-independent factor $\Phi$. A fixed time bin or bandwidth, a fixed bias ceiling, a finite temperature, a detector width or the intrinsic rate $\Gamma_0$ each introduce a second scale and break the exact rescaling. This is why condition (i) is needed, and why the numerical scan of \Sec{sec:test} tests the mean-current law rather than this statement: the interaction scan changes $\mu$ and $\delta/\mu$ as well as $\Xf$, and dividing by $(\mu\Xf)^2$ removes only the leading amplitude.

\textit{Homogeneous spectra.} Any homogeneous spectrum that satisfies the assumptions of \Sec{sec:single}, and for which the optimum in \Eq{eq:Cmax} is finite and positive, gives $\Fm^{\max}\propto a^2\Delta$, with a prefactor and optimum set by its shape. A sum of back-action channels of different degrees, $S(\pm\Delta;V)=\sum_ic_i(eV)^{r_i}s_i(\pm\Delta/eV)$, gives an asymmetry $\Asym$ that depends on $eV/\Delta$ alone whenever one channel dominates, so the degree of that channel enters only the prefactor and the optimum, provided that the measurement strength multiplying $\Asym^2$ in \Eq{eq:Fsingle} remains the electrical $S(0;V)\propto eV$; if the measurement strength itself had degree $r$, the exponent would be $r-2$. This is the case of a phonon channel of any single degree that dominates the populations while the electrical noise sets the imprecision. The statement about the full record does not extend to it, because the ratio of transition rates to measurement strength then depends on the gap. Two channels of different degree introduce a bias scale at which the optimised information crosses over between two inverse-gap regimes, and relaxation dominated by the detector is still required, since $\Gamma_0$ is not homogeneous in the bias.

\section{Checks of the two-level rate description}\label{app:checks}

The results of the main text use a two-level probe, rates of second order in the coupling without coherences, the mean detector current, and a non-interacting sensor dot. This appendix tests each of these choices at the operating points of \Tab{tab:matched}. The scripts and their outputs are in the archive.

\subsection{Higher states of the probe}\label{app:multilevel}

The rate model of \Sec{sec:rate} is generalised to the lowest $N$ singlets of the probe. The field-dressed states and energies are obtained by diagonalising $H(\varepsilon)=\mathrm{diag}(E_i)+\varepsilon X$ in the $N$-level space, with the full dipole and detector matrices of the solver (\Tab{tab:kernels} lists the elements to $|2\rangle$); every pair of dressed states is connected by the detector-induced rate $(\lambda M'_{ij})^2S(E'_i-E'_j)/\hbar$ at the dressed transition energy, plus the intrinsic rate $\Gamma_0(1+n_B)$ downward and $\Gamma_0n_B$ upward with the same $\Gamma_0$ for every transition; the stationary populations $p_i$ follow from the $N$-state master equation; the mean signal is $\sum_ip_iM'_{ii}$ and its field derivative is taken numerically, so that the dressed levels, the dressed matrix elements and the populations all vary with the field; and the telegraph noise is replaced by the zero-frequency noise of the $N$-state jump process, $S_{\mathrm{jump}}=2\,\delta\bm m^{T}(\Pi-W)^{-1}\mathrm{diag}(\bm p)\,\delta\bm m$, with $W$ the rate matrix, $\Pi=\bm p\bm1^{T}$ and $\delta\bm m$ the signal levels minus their mean, which reduces to $S_{\mathrm{tel}}$ for $N=2$. For $N=2$ the model reproduces \Eq{eq:Fmean} to $10^{-6}$. The levels used are 0, 0.181, 0.551 and 0.761~meV for the pair and 0, 0.181, 0.446 and 0.849~meV for the matched single electron, whose third level is lower.

\begin{table}[t]
\caption{Matched comparison through the point contact with $N=2$, 3 and 4 singlets in the rate equations (bias ceiling 1~mV, $T_1=16$~ns). Ratio: pair-to-single information rate; Sens.: pair sensitivity in V\,m$^{-1}$\,Hz$^{-1/2}$; $eV^{*}$: optimum bias in \textmu V; $p_f$ and $p_{>}$: populations of $|f\rangle$ and of the higher states at the optimum (at $\lambda=5$~\textmu eV the optimum is at the 1~mV ceiling, $p_f=0.04$ and $p_{>}=1.8$--$1.9\times10^{-3}$ for $N\ge3$). The sensor-dot values of \Tab{tab:matched} are unchanged to 0.4\% for every $N$ (the fourth level of the single electron is odd and dipole-coupled to its ground state), with $p_{>}=0$. The last rows give the fitted gap exponents of the point contact with the 50-orbital scan of \Sec{sec:test}.}
\label{tab:multilevel}
\centering
\footnotesize
\setlength{\tabcolsep}{3.5pt}
\begin{tabular}{clcccccccccc}
\toprule
 & & \multicolumn{2}{c}{$\lambda=5$~\textmu eV} & \multicolumn{4}{c}{$\lambda=20$~\textmu eV} & \multicolumn{4}{c}{$\lambda=50$~\textmu eV}\\
\cmidrule(lr){3-4}\cmidrule(lr){5-8}\cmidrule(lr){9-12}
$N$ & $T$ & Ratio & Sens. & Ratio & Sens. & $eV^{*}$ & $p_{>}$ & Ratio & Sens. & $eV^{*}$ & $p_f$\\
\midrule
2 & 10~mK & 2.79 & 0.50 & 2.48 & 0.183 & 456 & 0 & 2.90 & 0.084 & 232 & 0.05\\
2 & 100~mK & 2.79 & 0.52 & 2.47 & 0.188 & 465 & 0 & 2.87 & 0.088 & 250 & 0.07\\
3 & 10~mK & 2.72 & 0.50 & 2.11 & 0.174 & 534 & $3.5\times10^{-3}$ & 2.30 & 0.082 & 302 & 0.10\\
3 & 100~mK & 2.72 & 0.51 & 2.10 & 0.179 & 532 & $3.5\times10^{-3}$ & 2.28 & 0.085 & 320 & 0.12\\
4 & 10~mK & 2.72 & 0.50 & 2.09 & 0.172 & 548 & $3.9\times10^{-3}$ & 2.29 & 0.081 & 326 & 0.12\\
4 & 100~mK & 2.72 & 0.51 & 2.08 & 0.177 & 542 & $3.9\times10^{-3}$ & 2.27 & 0.084 & 343 & 0.13\\
\midrule
\multicolumn{12}{l}{Exponents, $\lambda=50$~\textmu eV, 10 / 100~mK: 1.22 / 1.15 ($N=2$), 1.21 / 1.14 ($N=3$), 1.19 / 1.11 ($N=4$)}\\
\multicolumn{12}{l}{Exponents, $\lambda=20$~\textmu eV, 10 / 100~mK: 1.67 / 1.61 ($N=2$), 1.64 / 1.58 ($N=3$), 1.61 / 1.56 ($N=4$)}\\
\multicolumn{12}{l}{Zero temperature, $\lambda=50$~\textmu eV: 1.02 ($N=2$ and 3)}\\
\bottomrule
\end{tabular}
\end{table}

\TAB{tab:multilevel} gives the result for $N=2$, 3 and 4. For the sensor dot nothing changes beyond 0.4\% in the pair-to-single ratio: biased below the gap it cannot reach the third level at zero temperature, and at 100~mK the thermal tails populate it to less than $10^{-16}$, so every sensor-dot number of the main text holds for every $N$. For the point contact the third level matters in two ways. At $\lambda=5$ and 20~\textmu eV, where the optimum bias is 0.5--1~mV, it is populated to $2$--$4\times10^{-3}$ by sequential excitation from $|f\rangle$ and by direct excitation. This population reduces the signal, but the field mixing described next adds more than it removes, so the net effect, at fixed settings and after re-optimisation alike, is an improvement of the best sensitivity by 1--5\% (0.503 to 0.499, 0.183 to 0.174 and 0.084 to 0.082~V\,m$^{-1}$\,Hz$^{-1/2}$ at 10~mK for $\lambda=5$, 20 and 50~\textmu eV). At every coupling the field also mixes $|f\rangle$ with $|2\rangle$ ($X_{f2}=5.5\,a^{*}$, $E_2-E_f=0.37$~meV), which adds to the signal carried by the excited state; this moves the optimum bias up by 15--30\% with three levels and by up to 40\% with four and, because the single electron's third level is lower and more strongly coupled, lowers the pair-to-single ratio from 2.5--2.9 to 2.1--2.7, by up to 21\%. The fourth level changes the ratios and sensitivities of the $N=3$ model by about 1\% and the optimum bias by up to 8\%. The gap exponents of the point contact move by at most 0.06 and the zero-temperature exponent not at all, so the inverse-gap law of \Sec{sec:single} is unaffected. The bias sweep of \Fig{fig:bias} with $N=3$ is shown there as a dotted line.

The truncation is tested by extending the same rate model to six and eight singlets (the fifth to eighth lie at 0.98, 1.19, 1.27 and 1.48~meV for the pair and at 1.34, 1.96, 2.67 and 3.52~meV for the single electron, so a bias ceiling of 1~mV admits direct excitation of the fifth level and sequential excitation of several). At the optimised point-contact operating points of \Tab{tab:matched} the change from four to eight levels is below 1\% in the sensitivities (0.497 to 0.496, 0.172 to 0.171 and 0.0813 to 0.0810~V\,m$^{-1}$\,Hz$^{-1/2}$ at 10~mK for $\lambda=5$, 20 and 50~\textmu eV), below 1\% in the pair-to-single ratios (2.72 to 2.74, 2.09 to 2.11 and 2.29 to 2.30) and at most 1.4\% in the optimum bias, with the population of the higher levels at most $4\times10^{-3}$; the sensor dot with eight levels gives the ratio 3.00. This is a truncation test within the rate model at these operating points, not a global convergence certificate.

\subsection{Information in the full record}\label{app:fullrecord}

The mean-current information rate of \Eq{eq:Fmean} is a lower bound on the Fisher information of the full record. We have computed the latter numerically for the observation model of \App{app:detectors}: the probe jumps between $|0'\rangle$ and $|f'\rangle$ with the rates of \Eq{eq:rates}, and the record is $\mathrm{d}Y=m_s\,\mathrm{d}t+\sqrt{S_q}\,\mathrm{d}B$ with $m_{0,f}=M_{00}-a\varepsilon$, $M_{ff}+a\varepsilon$ and the field also entering the rates through $M_{0'f'}$. Records are simulated on a time step one hundredth of the shorter of the switching and measurement times, with jumps drawn with probability rate$\times\mathrm{d}t$; the log-likelihood of each record is propagated by the forward filter of the two-state hidden Markov process with the exact finite-step transition matrix, its derivative with respect to the field is obtained by running the filter at $\varepsilon=\pm h$ ($h=10^{-3}\Delta/\Xf$) on the same record, and the Fisher information rate is the variance of this score divided by the record length. This is a discretised estimate of the continuous-time quantity, with a statistical error, not an exact calculation. The estimator was checked in two limits, with the outputs in the archive: with fast switching ($\Gmeas/(\Gup+\Gdn)=0.013$) it gives $0.27\pm0.01$ in units of $a^2/S_q$, above $F_{\mathrm{mean}}=0.24$ as the data-processing inequality requires (the small excess is the information carried by the field dependence of the telegraph-noise amplitude, a channel the mean current does not have, which is negligible at the operating points of the paper where $S_{\mathrm{tel}}/S_q<10^{-2}$), and with resolved switching ($\Gmeas/(\Gup+\Gdn)=6$) it gives 0.89--0.96 against the resolved-path value 1 and $F_{\mathrm{mean}}=0.013$.

At the four point-contact operating points of \Tab{tab:matched} at $\lambda=20$ and 50~\textmu eV, the estimate of the ratio $\Phi=F_{\mathrm{full}}/\Fm$, averaged over six independent seeds of 8000 records of 400 switching times each, is $1.003\pm0.006$, $1.005\pm0.006$, $1.002\pm0.006$ and $1.005\pm0.006$ (standard error over seeds; a single seed has a Monte-Carlo error of about 0.016, and single-seed estimates scatter between 0.99 and 1.03). Halving or doubling the time step, the record length and $h$ changes the estimate at the first point by less than 0.02, within the statistical error; the refined runs sit slightly below 1, which the true value cannot, so the estimator carries a statistical and discretisation uncertainty of 1--2\%. The result is therefore that no excess information in the full record is resolved at these points within 1--2\%; it is not a measured bound on the excess, and the bound of \Sec{sec:rate} on resolving jumps (the measurement rate is only 0.5\% of the switching rate) is what makes a small excess expected. A rigorous upper bound follows from a record whose state path is known exactly: its information rate is $a^2/S_q$ from the two levels plus the information in the jump times, $[\Gup\Gdn/(\Gup+\Gdn)]\,[(\partial_\varepsilon\ln\Gup)^2+(\partial_\varepsilon\ln\Gdn)^2]$, where $\Gup\Gdn/(\Gup+\Gdn)$ is the stationary rate of jumps in each direction and the derivatives follow analytically from the dependence of the rates on $M_{0'f'}$. At the point-contact points this bound is 1.2--1.7 times $\Fm$ and is far from attained. For the sensor dot the excited population at the optimum is below $3\times10^{-7}$ and no trajectories are needed: the bound gives $\Phi-1<6.3\times10^{-7}$ at $\lambda=20$~\textmu eV and $<1.1\times10^{-6}$ at $\lambda=50$~\textmu eV at the working point and 100~mK. The same bound evaluated along the basis-extrapolated scan at 100~mK with $\Gamma=\lambda=20$~\textmu eV grows with the thermal excitation as the gap closes: $\Phi-1<4\times10^{-7}$, $1.6\times10^{-5}$ and $2.7\times10^{-4}$ at the gaps 186, 154 and 130~\textmu eV above the floor, and $2\times10^{-2}$ to $1.4\times10^{-1}$ at the points below it, where the probe is thermally excited and the law no longer holds anyway. At 10~mK the thermal tail $\mathrm{e}^{-(\Delta-eV)/k_BT}$ keeps the excited population below $2\times10^{-14}$ at every point of the scan and the bound below $10^{-13}$. These numbers are bounds on, not measurements of, the actual excess; for the sensor dot above the floor the full record and the mean current carry the same information to better than $3\times10^{-4}$ at 100~mK and to numerical precision at 10~mK, and for the point contact the full-record statement is confined to the four points tested.

\subsection{Coherences and static renormalisation}\label{app:coherences}

The rates of \Eq{eq:rates} are the secular limit of the Bloch--Redfield equation for the probe coupled to the detector through $H_{\mathrm{int}}=\lambda MQ$. We have solved the Redfield equation for the lowest two and three singlets at five of the operating points of \Tab{tab:matched} (listed in \Sec{sec:limits}), using the same spectra $S(E)$ as in the main text and the same intrinsic relaxation on every transition. The dissipative part is kept in full, without the secular approximation: $\dot\rho=-\mathrm{i}[H+H_{LS},\rho]+\sum_{(ab),(cd)}\tfrac12\lambda^2S(\omega_{ab})[A_{ab}\rho A_{cd}^\dagger-A_{cd}^\dagger A_{ab}\rho]+\mathrm{h.c.}$, with $A_{ab}=M'_{ab}|a\rangle\langle b|$ in the field-dressed basis and $\omega_{ab}=E'_b-E'_a$, so that the coupling between populations and coherences and the contribution $2\mu\,\mathrm{Re}\,\rho_{0f}$ of the coherence to the mean signal are retained. The Lamb shift is kept in its secular, Hermitian form, $H_{LS}=\sum_{(ab)}S_L(\omega_{ab})A_{ab}^\dagger A_{ab}$ with $S_L(\omega)=(2\pi)^{-1}\,\mathcal{P}\!\int\mathrm{d}\omega'\lambda^2S(\omega')/(\omega-\omega')$, added to the coherent part of the generator while the dissipative spectra are evaluated at the unshifted dressed energies. The model is therefore a non-secular dissipative Redfield equation with a secular Lamb-shift Hamiltonian. The quantities compared are the stationary populations, the static coherence and the field derivative of the mean signal, which is the numerator of \Eq{eq:Fmean}; the noise denominator is not recomputed, so this is a check of the populations and the signal, not of the full information rate.

The results are as follows. Without the Lamb shift, the stationary excited population (in absolute terms) and the signal derivative $\partial_\varepsilon\bar M$ (in relative terms) agree with the rate model to better than $10^{-7}$ at every operating point, with two and with three levels, and the static coherence $|\rho_{0f}|$ is below $4\times10^{-6}$. With the secular Lamb shift the gap is lowered by $1.0$~\textmu eV for the sensor dot at $\lambda=\Gamma=20$~\textmu eV and 10~mK, $0.4$~\textmu eV at 100~mK and $6.3$~\textmu eV at $\lambda=50$~\textmu eV, up to 3.5\% of the gap, and by $0.1$--$0.6$~\textmu eV for the point contact (a cutoff-dependent value, since its spectrum grows linearly at large energy; the cutoff is 2~meV); the populations and the signal derivative still agree with the rate model to $10^{-7}$, because the shift is diagonal in the dressed basis and small compared with the gap. We have not kept the non-secular cross terms of the Lamb shift between components of different frequency. When they are kept literally, the stationary excited population of the sensor dot at zero temperature becomes of order $10^{-1}$ and grows roughly in proportion to the Lamb shift; we interpret this as the non-secular Redfield equation ceasing to give a controlled stationary state when the reactive couplings exceed the dissipative rates that would balance them, which is the situation in the energy-filtered regime, $S(\pm\Delta)\ll S(0)$, but we have not diagnosed it independently, and the full non-secular reactive treatment is not validated here. Dropping these terms is a choice, and the agreement reported is a statement about the specified model, non-secular dissipative with a secular Lamb shift, at these points, for the populations and the signal; it does not establish the accuracy of a fully recalibrated detector--probe model, and it does not bound higher orders in the coupling, detector memory, or correlations between the output noise and the back-action.

\subsection{Coulomb blockade in the sensor dot}\label{app:blockade}

The resonant-level formulas of \App{app:detectors} are exact at any temperature for a non-interacting level in the wide-band limit; what they omit is the charging energy. A Coulomb-blockaded dot with charging energy $U$ larger than the bias, the gap and the level width is read through one charge transition, and in the sequential-tunnelling regime $\Gamma\ll k_BT$ its transport is described by the occupation master equation with energy-resolved lead rates, into which the charging energy and the spin degeneracy enter: lead rates $\Gamma_\alpha f_\alpha(\epsilon_0)$ in and $\Gamma_\alpha[1-f_\alpha(\epsilon_0)]$ out ($\alpha=L,R$, $\Gamma_L=\Gamma_R=\Gamma/2$), with a factor 2 on the rate in for a spin-degenerate level with $U\to\infty$. The current and its zero-frequency noise follow from the counting statistics of this two-state process, the gain is $\partial I/\partial\epsilon_0$, and the dephasing uses the telegraph charge noise $2n(1-n)/(\Gamma_{\mathrm{in}}+\Gamma_{\mathrm{out}})$. A probe transition of energy $\pm\Delta$ that accompanies a tunnelling event needs a lead electron at $\epsilon_0\pm\Delta$, so the excitation and relaxation weights seen by the probe are $S(\mp\Delta)=\Delta^{-2}\sum_\alpha\Gamma_\alpha[f_\alpha(\epsilon_0\pm\Delta)(1-n)s+(1-f_\alpha(\epsilon_0\mp\Delta))n]$ with $s=2$ for the spinful case, which is the $\Gamma\ll k_BT$ limit of the resonant-level spectrum and vanishes at zero temperature for $eV<\Delta$: energy filtering rests on the lead distributions, not on the coherence of the level. As a check of the implementation, the spinless sequential dot and the resonant level agree in gain, noise, measurement rate and relaxation to 4\% at $\Gamma/k_BT=0.06$ and to 14\% at $\Gamma/k_BT=0.23$ (100~mK, $\Gamma=0.5$ and 2~\textmu eV); this checks the spinless limit only, and the spin-degenerate rates are the standard ones of sequential tunnelling.

Optimised at the working gap with $\lambda=20$~\textmu eV, the best sensitivities (in V\,m$^{-1}$\,Hz$^{-1/2}$) are 0.108 for the resonant level, 0.101 for the spinless sequential dot and 0.086 for the spin-degenerate blockaded dot at 100~mK and $\Gamma=2$~\textmu eV ($\Gamma/k_BT=0.23$); 0.057, 0.051 and 0.043 at 50~mK and $\Gamma=2$~\textmu eV ($\Gamma/k_BT=0.46$, a marginal case for the sequential description); and 0.105, 0.101 and 0.086 at 50~mK and $\Gamma=0.5$~\textmu eV ($\Gamma/k_BT=0.12$). In every case the optimum bias lies below the gap, at 40--140~\textmu V against $\Delta=181$~\textmu eV, and the excited population is of order $10^{-4}$ or below. With the level re-optimised at each bias, the measurement rate saturates for $eV\gtrsim5k_BT$ (40~\textmu V at 100~mK) at $1.1$--$1.6\times10^{6}$~s$^{-1}$ for the sequential dots against $1.0\times10^{6}$~s$^{-1}$ for the resonant level of the same width. Coulomb blockade therefore does not suppress the gain below the gap in this regime; it changes the scale on which the gain saturates from $\Gamma$ to a few $k_BT$, and the spin-degenerate dot is 18--25\% more sensitive than the non-interacting level of the same width because its Coulomb peak is steeper. The inverse-square law is then expected to hold above the thermal floor by the argument of \Sec{sec:filtered}, since its two ingredients, full measurement strength below the gap and no emission at the gap, are present; we have not measured the exponent for the sequential dot with a gap scan. What the sequential regime costs is the absolute measurement rate, which falls as $\Gamma/k_BT$: relative to the $\Gamma=20$~\textmu eV level of the main text (0.053 at 100~mK) the $\Gamma=2$~\textmu eV dots lose a factor of two in sensitivity, and their efficiency $\eta$ is far below the quantum limit ($\eta=0.0005$--0.01 here, as for the resonant level of the same width), because the gain is set by $k_BT$ while the noise is not; an incoherent detector is in addition limited as in~\cite{Clerk2004}. These examples show that energy filtering and operation below the gap survive Coulomb blockade in the sequential-tunnelling regime with $U\gg\max(eV,\Delta,\Gamma,k_BT)$ and $|\lambda\delta|\ll U$, as for the sensor dots of~\cite{Barthel2010,Haldar2025} with $U$ of order 1~meV. They do not establish that the non-interacting resonant level is quantitatively accurate for a spin-degenerate interacting dot at $\Gamma\gtrsim k_BT$, which remains an assumption.

\bibliographystyle{iopnum}
\bibliography{references}

\begin{thebibliography}{99}
\small\setlength{\itemsep}{0pt}

\bibitem{Zanardi2008} Zanardi P, Paris M~G~A and {Campos Venuti} L 2008 Quantum
  criticality as a resource for quantum estimation
  \href{https://doi.org/10.1103/PhysRevA.78.042105}{\textit{Phys. Rev. A}
  \textbf{78} 042105}

\bibitem{Rams2018} Rams M~M, Sierant P, Dutta O, Horodecki P and Zakrzewski J
  2018 At the limits of criticality-based quantum metrology: apparent
  super-Heisenberg scaling revisited
  \href{https://doi.org/10.1103/PhysRevX.8.021022}{\textit{Phys. Rev. X}
  \textbf{8} 021022}

\bibitem{Garbe2020} Garbe L, Bina M, Keller A, Paris M~G~A and Felicetti S 2020
  Critical quantum metrology with a finite-component quantum phase transition
  \href{https://doi.org/10.1103/PhysRevLett.124.120504}{\textit{Phys. Rev.
  Lett.} \textbf{124} 120504}

\bibitem{Montenegro2025} Montenegro V, Mukhopadhyay C, Yousefjani R, Sarkar S,
  Mishra U, Paris M~G~A and Bayat A 2025 Review: quantum metrology and sensing
  with many-body systems
  \href{https://doi.org/10.1016/j.physrep.2025.05.005}{\textit{Phys. Rep.}
  \textbf{1134} 1--62}

\bibitem{Mihailescu2026tut} Mihailescu G, Alushi U, {Di Candia} R, Felicetti S
  and Gietka K 2026 Critical quantum sensing: a tutorial on parameter
  estimation near quantum phase transitions
  \href{https://doi.org/10.1103/v7mf-yh8n}{\textit{PRX Quantum} \textbf{7}
  020201}

\bibitem{Mihailescu2024nano} Mihailescu G, Kiely A and Mitchell A~K 2024
  Quantum sensing with nanoelectronics: Fisher information for an applied
  perturbation \href{https://arxiv.org/abs/2406.18662}{arXiv:2406.18662}

\bibitem{Abdullah2009} Abdullah S, Coe J~P and D'Amico I 2009 Effect of
  confinement potential geometry on entanglement in quantum dot-based
  nanostructures
  \href{https://doi.org/10.1103/PhysRevB.80.235302}{\textit{Phys. Rev. B}
  \textbf{80} 235302}

\bibitem{Coe2010} Coe J~P, Abdullah S and D'Amico I 2010 Geometry induced
  entanglement transitions in nanostructures
  \href{https://doi.org/10.1063/1.3364065}{\textit{J. Appl. Phys.} \textbf{107}
  09E110}

\bibitem{Gietka2022} Gietka K, Ruks L and Busch T 2022 Understanding and
  improving critical metrology. Quenching superradiant light-matter systems
  beyond the critical point
  \href{https://doi.org/10.22331/q-2022-04-27-700}{\textit{Quantum} \textbf{6}
  700}

\bibitem{Gyhm2026} Gyhm {\relax J-Y}, Kwon H and Hwang {\relax M-J} 2026
  Fundamental scaling limit in critical quantum metrology
  \href{https://doi.org/10.1103/xwlk-2zfp}{\textit{Phys. Rev. Res.} \textbf{8}
  033004}

\bibitem{Mihailescu2026gap} Mihailescu G and Gietka K 2026 Mind the gap:
  anti-critical quantum metrology
  \href{https://arxiv.org/abs/2602.03675}{arXiv:2602.03675}

\bibitem{Mihailescu2024qst} Mihailescu G, Bayat A, Campbell S and Mitchell A~K
  2024 Multiparameter critical quantum metrology with impurity probes
  \href{https://doi.org/10.1088/2058-9565/ad438d}{\textit{Quantum Sci.
  Technol.} \textbf{9} 035033}

\bibitem{Mihailescu2025pra} Mihailescu G, Campbell S and Gietka K 2025
  Uncertain quantum critical metrology: from single- to multiparameter sensing
  \href{https://doi.org/10.1103/PhysRevA.111.052621}{\textit{Phys. Rev. A}
  \textbf{111} 052621}

\bibitem{Korotkov1999} Korotkov A~N 1999 Continuous quantum measurement of a
  double dot \href{https://doi.org/10.1103/PhysRevB.60.5737}{\textit{Phys. Rev.
  B} \textbf{60} 5737}

\bibitem{Korotkov2001} Korotkov A~N 2001 Selective quantum evolution of a qubit
  state due to continuous measurement
  \href{https://doi.org/10.1103/PhysRevB.63.115403}{\textit{Phys. Rev. B}
  \textbf{63} 115403}

\bibitem{Gurvitz1997} Gurvitz S~A 1997 Measurements with a noninvasive detector
  and dephasing mechanism
  \href{https://doi.org/10.1103/PhysRevB.56.15215}{\textit{Phys. Rev. B}
  \textbf{56} 15215}

\bibitem{Goan2001} Goan {\relax H-S}, Milburn G~J, Wiseman H~M and Sun H~B 2001
  Continuous quantum measurement of two coupled quantum dots using a point
  contact: a quantum trajectory approach
  \href{https://doi.org/10.1103/PhysRevB.63.125326}{\textit{Phys. Rev. B}
  \textbf{63} 125326}

\bibitem{Clerk2010} Clerk A~A, Devoret M~H, Girvin S~M, Marquardt F and
  Schoelkopf R~J 2010 Introduction to quantum noise, measurement, and
  amplification \href{https://doi.org/10.1103/RevModPhys.82.1155}{\textit{Rev.
  Mod. Phys.} \textbf{82} 1155}

\bibitem{Young2010} Young C~E and Clerk A~A 2010 Inelastic backaction due to
  quantum point contact charge fluctuations
  \href{https://doi.org/10.1103/PhysRevLett.104.186803}{\textit{Phys. Rev.
  Lett.} \textbf{104} 186803}

\bibitem{Gasser2009} Gasser U, Gustavsson S, K\"ung B, Ensslin K, Ihn T,
  Driscoll D~C and Gossard A~C 2009 Statistical electron excitation in a double
  quantum dot induced by two independent quantum point contacts
  \href{https://doi.org/10.1103/PhysRevB.79.035303}{\textit{Phys. Rev. B}
  \textbf{79} 035303}

\bibitem{Harbusch2010} Harbusch D, Taubert D, Tranitz H~P, Wegscheider W and
  Ludwig S 2010 Phonon-mediated versus Coulombic backaction in quantum dot
  circuits \href{https://doi.org/10.1103/PhysRevLett.104.196801}{\textit{Phys.
  Rev. Lett.} \textbf{104} 196801}

\bibitem{Granger2012} Granger G \textit{et al} 2012 Quantum interference and
  phonon-mediated back-action in lateral quantum-dot circuits
  \href{https://doi.org/10.1038/nphys2326}{\textit{Nat. Phys.} \textbf{8} 522}

\bibitem{Gammelmark2013} Gammelmark S and M{\o}lmer K 2013 Bayesian parameter
  inference from continuously monitored quantum systems
  \href{https://doi.org/10.1103/PhysRevA.87.032115}{\textit{Phys. Rev. A}
  \textbf{87} 032115}

\bibitem{Gammelmark2014} Gammelmark S and M{\o}lmer K 2014 Fisher information
  and the quantum Cram\'er--Rao sensitivity limit of continuous measurements
  \href{https://doi.org/10.1103/PhysRevLett.112.170401}{\textit{Phys. Rev.
  Lett.} \textbf{112} 170401}

\bibitem{Kiilerich2016} Kiilerich A~H and M{\o}lmer K 2016 Bayesian parameter
  estimation by continuous homodyne detection
  \href{https://doi.org/10.1103/PhysRevA.94.032103}{\textit{Phys. Rev. A}
  \textbf{94} 032103}

\bibitem{Albarelli2018} Albarelli F, Rossi M~A~C, Tamascelli D and Genoni M~G
  2018 Restoring Heisenberg scaling in noisy quantum metrology by monitoring
  the environment
  \href{https://doi.org/10.22331/q-2018-12-03-110}{\textit{Quantum} \textbf{2}
  110}

\bibitem{Onac2006} Onac E, Balestro F, {Willems van Beveren} L~H, Hartmann U,
  Nazarov Y~V and Kouwenhoven L~P 2006 Using a quantum dot as a high-frequency
  shot noise detector
  \href{https://doi.org/10.1103/PhysRevLett.96.176601}{\textit{Phys. Rev.
  Lett.} \textbf{96} 176601}

\bibitem{Gustavsson2007} Gustavsson S, Studer M, Leturcq R, Ihn T, Ensslin K,
  Driscoll D~C and Gossard A~C 2007 Frequency-selective single-photon detection
  using a double quantum dot
  \href{https://doi.org/10.1103/PhysRevLett.99.206804}{\textit{Phys. Rev.
  Lett.} \textbf{99} 206804}

\bibitem{Horibe2015} Horibe K, Kodera T and Oda S 2015 Back-action-induced
  excitation of electrons in a silicon quantum dot with a single-electron
  transistor charge sensor
  \href{https://doi.org/10.1063/1.4907894}{\textit{Appl. Phys. Lett.}
  \textbf{106} 053119}

\bibitem{Haldar2025} Haldar S, Munk M, Havir H, Khan W, Lehmann S, Thelander C,
  Dick K~A, Samuelsson P, Potts P~P and Maisi V~F 2025 Coherence of an
  electronic two-level system under continuous charge sensing by a quantum dot
  detector \href{https://doi.org/10.1103/PhysRevLett.134.023601}{\textit{Phys.
  Rev. Lett.} \textbf{134} 023601}

\bibitem{Schulenborg2014} Schulenborg J, Splettstoesser J, Governale M and
  Contreras-Pulido L~D 2014 Detection of the relaxation rates of an interacting
  quantum dot by a capacitively coupled sensor dot
  \href{https://doi.org/10.1103/PhysRevB.89.195305}{\textit{Phys. Rev. B}
  \textbf{89} 195305}

\bibitem{Sankar2025} Sankar S, Lotem M, Folk J, Sela E and Meir Y 2025
  Back-action effects in charge detection
  \href{https://doi.org/10.1103/mdsg-n9nk}{\textit{Phys. Rev. B} \textbf{112}
  075151}

\bibitem{Grant2026} Grant W \textit{et al} 2026 Anderson orthogonality as
  measurement backaction in coupled quantum dots
  \href{https://arxiv.org/abs/2608.06550}{arXiv:2608.06550}

\bibitem{Ilias2022} Ilias T, Yang D, Huelga S~F and Plenio M~B 2022
  Criticality-enhanced quantum sensing via continuous measurement
  \href{https://doi.org/10.1103/PRXQuantum.3.010354}{\textit{PRX Quantum}
  \textbf{3} 010354}

\bibitem{Ilias2024} Ilias T, Yang D, Huelga S~F and Plenio M~B 2024
  Criticality-enhanced electric field gradient sensor with single trapped ions
  \href{https://doi.org/10.1038/s41534-024-00833-w}{\textit{npj Quantum Inf.}
  \textbf{10} 36}

\bibitem{Yang2023} Yang D, Huelga S~F and Plenio M~B 2023 Efficient information
  retrieval for sensing via continuous measurement
  \href{https://doi.org/10.1103/PhysRevX.13.031012}{\textit{Phys. Rev. X}
  \textbf{13} 031012}

\bibitem{Zhang2025} Zhang C, Cirio M, Li {\relax X-Q} and Liang P 2025
  Enhancing information retrieval in quantum-optical critical systems via
  quantum measurement backaction
  \href{https://arxiv.org/abs/2511.22248}{arXiv:2511.22248}

\bibitem{OConnor2024} O'Connor E, Campbell S and Landi G~T 2024 Fisher
  information rates in sequentially measured quantum systems
  \href{https://doi.org/10.1088/1367-2630/ad345c}{\textit{New J. Phys.}
  \textbf{26} 033048}

\bibitem{Field1993} Field M, Smith C~G, Pepper M, Ritchie D~A, Frost J~E~F,
  Jones G~A~C and Hasko D~G 1993 Measurements of Coulomb blockade with a
  noninvasive voltage probe
  \href{https://doi.org/10.1103/PhysRevLett.70.1311}{\textit{Phys. Rev. Lett.}
  \textbf{70} 1311}

\bibitem{Cassidy2007} Cassidy M~C, Dzurak A~S, Clark R~G, Petersson K~D, Farrer
  I, Ritchie D~A and Smith C~G 2007 Single shot charge detection using a
  radio-frequency quantum point contact
  \href{https://doi.org/10.1063/1.2809370}{\textit{Appl. Phys. Lett.}
  \textbf{91} 222104}

\bibitem{Reilly2007} Reilly D~J, Marcus C~M, Hanson M~P and Gossard A~C 2007
  Fast single-charge sensing with a rf quantum point contact
  \href{https://doi.org/10.1063/1.2794995}{\textit{Appl. Phys. Lett.}
  \textbf{91} 162101}

\bibitem{Barthel2010} Barthel C, Kj{\ae}rgaard M, Medford J, Stopa M, Marcus
  C~M, Hanson M~P and Gossard A~C 2010 Fast sensing of double-dot charge
  arrangement and spin state with a radio-frequency sensor quantum dot
  \href{https://doi.org/10.1103/PhysRevB.81.161308}{\textit{Phys. Rev. B}
  \textbf{81} 161308}

\bibitem{Lieb1962} Lieb E and Mattis D 1962 Theory of ferromagnetism and the
  ordering of electronic energy levels
  \href{https://doi.org/10.1103/PhysRev.125.164}{\textit{Phys. Rev.}
  \textbf{125} 164}

\bibitem{Petta2004} Petta J~R, Johnson A~C, Marcus C~M, Hanson M~P and Gossard
  A~C 2004 Manipulation of a single charge in a double quantum dot
  \href{https://doi.org/10.1103/PhysRevLett.93.186802}{\textit{Phys. Rev.
  Lett.} \textbf{93} 186802}

\bibitem{Aguado2000} Aguado R and Kouwenhoven L~P 2000 Double quantum dots as
  detectors of high-frequency quantum noise in mesoscopic conductors
  \href{https://doi.org/10.1103/PhysRevLett.84.1986}{\textit{Phys. Rev. Lett.}
  \textbf{84} 1986}

\bibitem{Khym2006} Khym G~L, Lee Y and Kang K 2006 Controlled dephasing and
  measurement efficiency via quantum dot detector
  \href{https://doi.org/10.1143/JPSJ.75.063707}{\textit{J. Phys. Soc. Japan}
  \textbf{75} 063707}

\bibitem{Clerk2003} Clerk A~A, Girvin S~M and Stone A~D 2003 Quantum-limited
  measurement and information in mesoscopic detectors
  \href{https://doi.org/10.1103/PhysRevB.67.165324}{\textit{Phys. Rev. B}
  \textbf{67} 165324}

\bibitem{Pilgram2002} Pilgram S and B\"uttiker M 2002 Efficiency of mesoscopic
  detectors \href{https://doi.org/10.1103/PhysRevLett.89.200401}{\textit{Phys.
  Rev. Lett.} \textbf{89} 200401}

\bibitem{Ouyang2010} Ouyang {\relax S-H}, Lam {\relax C-H} and You J~Q 2010
  Backaction of a charge detector on a double quantum dot
  \href{https://doi.org/10.1103/PhysRevB.81.075301}{\textit{Phys. Rev. B}
  \textbf{81} 075301}

\bibitem{Xu2014} Xu F, Holmqvist C and Belzig W 2014 Overbias light emission
  due to higher-order quantum noise in a tunnel junction
  \href{https://doi.org/10.1103/PhysRevLett.113.066801}{\textit{Phys. Rev.
  Lett.} \textbf{113} 066801}

\bibitem{Stace2004} Stace T~M and Barrett S~D 2004 Continuous quantum
  measurement: inelastic tunneling and lack of current oscillations
  \href{https://doi.org/10.1103/PhysRevLett.92.136802}{\textit{Phys. Rev.
  Lett.} \textbf{92} 136802}

\bibitem{Stace2003} Stace T~M and Barrett S~D 2003 Continuous measurement of a
  charge qubit with a point contact detector at arbitrary bias: the role of
  inelastic tunnelling
  \href{https://arxiv.org/abs/cond-mat/0309610}{arXiv:cond-mat/0309610}

\bibitem{Clerk2004} Clerk A~A and Stone A~D 2004 Noise and measurement
  efficiency of a partially coherent mesoscopic detector
  \href{https://doi.org/10.1103/PhysRevB.69.245303}{\textit{Phys. Rev. B}
  \textbf{69} 245303}

\end{thebibliography}

\end{document}